\documentclass[aps,showpacs,prd,twocolumn]{revtex4-1}
\usepackage{graphicx}
\usepackage{amssymb}
\usepackage{amsmath}
\usepackage[dvipsnames]{xcolor}
\usepackage{float}
\usepackage{tabularx}
\usepackage{ulem}
\usepackage{accents}
\usepackage{graphicx}
\usepackage{amsfonts}
\usepackage{wasysym}
\usepackage[toc,page]{appendix}
\usepackage{verbatim}
\usepackage{booktabs}
\newcommand{\be}{\begin{equation}}
\newcommand{\ee}{\end{equation}}
\newcommand{\bq}{\begin{eqnarray}}
\newcommand{\eq}{\end{eqnarray}}

\begin{document}

\title{Biphoton phase-space correlations from Gouy-phase measurements using  double slits}

\author{F. C. V. de Brito  $^{a,b,c}$}\email[]{crislane.brito@ufabc.edu.br}
\author{I. G. da Paz$^d$}\email[]{irismarpaz@ufpi.edu.br}
\author{J. B. Araujo $^e$}
\author{Marcos Sampaio $^a$}

\affiliation{$^{a}$  CCNH, Universidade Federal do ABC,  09210-580 , Santo Andr\'e - SP, Brazil}

\affiliation{$^{b}$ Institute for Quantum Optics and Quantum Information—IQOQI Vienna, Austrian Academy of Sciences, Boltzmanngasse 3, 1090 Vienna, Austria}

\affiliation{$^{c}$ Vienna Center for Quantum Science and Technology, Atominstitut, TU Wien, 1020 Vienna, Austria}

\affiliation{$^d$ Universidade Federal do Piau\'{\i}, Departamento de F\'{\i}sica, 64049-550, Teresina - PI, Brazil}

\affiliation{$^e$ ColibrITD - Quantum Computing Group, 1-7 Cours de Valmy, Puteaux 92800, France.}

\begin{abstract}
\noindent
Quantum correlations encoded in photonic Laguerre-Gaussian modes were shown to be related to the Gouy phase shifts (D. Kawase et al., Phys. Rev. Lett. 101, 050501 (2008))
allowing for a non-destructive manipulation of photonic quantum states. In this work we exploit the relation between phase space correlations of biphotons produced by spontaneously parametric down conversion (SPDC) as encoded in the logarithmic negativity (LN) and the Gouy phase as they are diffracted through an asymmetrical double slit setup. Using an analytical approach based on a double-Gaussian approximation for type-I SPDC biphotons, we show that measurements of Gouy phase differences provide information on their phase space entanglement variation, governed by the physical parameters of the experiment and expressed by the LN via covariance matrix elements.

\end{abstract}
\maketitle
\noindent

\section{Introduction}
\label{SectionI}

Quantum information science has reached a revolutionary point in the timeline of technological advances and commercial
exploitation. 
Quantum effects have been effectively used in various kinds of information processing and communication. 
More recently, continuous quantum variables \cite{ADESSO} have been used to extend quantum communication protocols from discrete (finite dimension) to continuous variables (infinite dimensions). This amounts to encode information in continuous variables, such as quadratures of coherent states. 
The main advantage of working with continuous variables is their unconditionalness \cite{BRAUNSTEIN}, meaning that the information carriers (e.g. entangled states) are originated from the nonlinear optical interaction of a laser with a crystal  in an unconditional (every inverse bandwidth time). 
The drawback is the quality of the entanglement of the states. Quantum information via continuous variables  has been contemplated in quantum key distribution protocols \citep{QKD}, in which one may convey unconditionally secure communication. Continuum variable quantum correlations can be encoded in degrees of freedom with a continuous spectrum such as the position and momentum of a particle as well as light quadratures and collective magnetic moments of atomic ensembles, which obey the same canonical algebra. 
In this contribution we study the quantification of quantum correlations encoded in continuous variables of biphotons produced in a nonlinear crystal via spontaneous parametric down-conversion as encoded in their Gouy phase in a double slit experiment.

In a nutshell, the Gouy phase is an axial phase shift that a converging light wave suffers as it passes through a focal point. This phase anomaly was first observed by L. G. Gouy in 1890 \cite{gouy1,gouy2} and it can be seen as the result of the photon transverse momentum spread as its trajectory is limited in the transverse direction due to focusing  or diffracting  through a slit aperture. 
However, the Gouy phase shift is present in any kind of wave that is transversally confined, and its physical interpretation is connected to the underlying wave phenomena \cite{Visser1,simon1993,feng2001,yang,boyd,hariharan,feng98, Pang}. For instance, the Gouy phase shift can be seen as a manifestation
of a general Berry phase, namely a topological phase acquired by a system after a cyclic adiabatic evolution in parameter space \citep{BERRY}. 
It is known that the Gouy phase shift can be understood in light of the uncertainty principle, as transverse spatial confinement leads to a spread in the transverse momenta and, consequently, to a shift in the expectation value of the axial propagation constant. 
Thus,  a general expression for the Gouy phase shift is given in terms of expectation values of the squares of the transverse momenta. Therefore, the physical origin of the Gouy phase shift arises from the covariance matrix elements \citep{feng2001, hariharan}.
The Gouy phase shift amounts to a  $n \times \pi/2$  axial
phase shift that a converging light wave experiences
as it passes through its focus in propagating from $-\infty$ to $+\infty$.
Whereas the Gouy phase shift for a focused wave  is equal to $\pi/2$ for cylindrical waves (line focus) and $\pi$ for spherical waves (point focus),  
in wave diffraction by small apertures it was verified to be $\pi/4$ \cite{Paz4}. 

Gouy phase shift is observed in different wave phenomena, such as water waves \cite{chauvat}, acoustic \cite{holme}, surface plasmon-polariton \cite{zhu}, phonon-polariton \cite{feurer} pulses, and more recently in matter waves \cite{cond,elec2,elec1} and it has important applications in modern optics and photonics. 
For example, the Gouy phase was used to determine the resonant frequencies in laser cavities \cite{siegman}, the phase matching in high-order harmonic generation (HHG) \cite{Balcou} as well as to describe  the spatial variation of the carrier envelope phase of ultra-short pulses in a laser focus \cite{Lindner}. 
Moreover, it has been observed that the Gouy phase influences the evolution of optical \cite{Allen} and electronical vortex beams, which acquire an additional Gouy phase dependence on the absolute value of the orbital angular momentum \cite{elec2}.
For coherent matter waves, the Gouy phase has been studied in \cite{Paz4,Paz1,Paz2,Ducharme} and experimental realizations were performed in different systems such as Bose-Einstein condensates \cite{cond}, electron vortex beams \cite{elec2} and astigmatic electron matter waves \cite{elec1}. 
Matter wave Gouy phase shifts can be used as mode converters in quantum information processing \cite{Paz1}, in the development of singular electron optics \cite{elec1} and in the study of non-classical looped path contributions in multiple slit interferometry \cite{Paz3}.

In our analysis, we use entangled photon pairs (biphotons or twin-photons) produced via spontaneous parametric down conversion (SPDC). 
Such events are produced when a nonlinear crystal is hit by a  (pump) photon with frequency $\omega_p$, which in turn produces two new outgoing photons of lower frequencies $\omega_s$ (signal) and $\omega_i$ (idler).  
A type-I SPDC process happens when the polarization of the outgoing pair of photons is parallel to each other and orthogonal to the polarization of the pump photon. 
The spatial distribution of the emerging photons forms a cone that is aligned with the pump beam propagation, with the apex at the crystal. 
Such a process is energy-momentum conserving, namely $\omega_p=\omega_i + \omega_s$ and $\vec{k}_p= \vec{k}_i + \vec{k}_s$, and thus the outgoing photon pair state is highly correlated in their spatial, temporal, spectral and polarization properties \cite{Yanhua}. 
Therefore, their joint quantum state is entangled \cite{ENTANGBIPH}.
Under reasonable physical assumptions, a double-Gaussian effective wavefunction for a type-I SPDC can be constructed from the quantization of the non-linear interaction in the medium represented by the crystal which allows for two outgoing photons from the pump beam source which may be treated classically \cite{howell}-\cite{Wolf1995}. 
We study the propagation of the twin photons diffracted by a double-slit setup in order to obtain the wavefunctions corresponding to the four possible paths, namely both photons passing through the upper (lower) slit, or each one passing through a different slit. 
The purpose here is to generalize the results cast in  \cite{crisfree}, where we established a connection between the logarithmic negativity and the Gouy phase in the biphoton free propagation, as both depend on the biphoton correlations through the Rayleigh length. 
Furthermore, by focusing the double Gaussian biphoton wavefunction using a thin lens, we have calculated the Gouy phase by writing the quantities as a function of the Rayleigh range and have found good agreement with the experimental data \cite{Kawase}.

In this contribution, we calculate the Gouy phase and the logarithmic negativity at the detection screen after the diffraction of a type-I SPDC biphoton wavepacket. 
In contrast to the free evolution, in which the biphoton entanglement remains constant, a diffraction through the double-slit changes the phase space correlations \cite{Pham}. We analyze the quantum correlations' behavior of twin photons diffracting through a double-slit, and from these correlations presented in the covariance matrix, one can calculate an entanglement quantifier known as logarithmic negativity.
The biphoton entanglement at the detection screen depends on the double-slit geometrical parameters as well as on the initial wavepacket Gaussian widths. 
We show that the logarithmic negativity and the Gouy phase difference are connected to each other through the slit width, thus, this setup can serve the purpose of measuring the variation of entanglement, produced by the change in spatial geometry, in terms of the slit width using indirect measurements of the Gouy phase shift.
The Gouy phase difference is experimentally accessible through the relative intensity and visibility for certain values of slit widths in an asymmetrical double-slit.
We follow the ideas presented in \cite{Kawase, crisfree} where it was shown that biphoton phase space correlations are related with the Gouy phase via the logarithmic negativity. 
For this purpose, we use an asymmetrical double-slit (different slit widths) in order to generate a Gouy phase shift, since this phase cancels out when the slits have the same widths.

We organize our results as follows. In section II, we
study the time evolution of a type-I SPDC biphoton wavefunction diffracted through an asymmetrical double-slit and obtain the corresponding Gouy phase. 
Then, we calculate the covariance matrix elements and the biphoton entanglement through the logarithmic negativity at the
detection screen in terms of the slit parameters. 
The biphoton position cross-correlations both at the slits and at the detection screen are computed for biphotons passing through the same slit as well as through different slits as a function of the geometrical parameters and the wavepacket Gaussian widths. 
We verify that for a specific set of those parameters,  the interference pattern can be fitted by the interference of the wavefunctions corresponding to the photons passing through the same slit only. 
For this case, we define the fringe visibility in analogy to the double-slit single particle interferometry and study the behavior of the visibility as a function of the logarithmic negativity.
Section III is devoted to the study of the Gouy phase difference for an asymmetrical double-slit.
We show that the Gouy phase difference can be expressed in terms of the relative intensity and the fringe visibility, which can be measured in a simple way. 
Then we fix the double-slit parameters as well as the propagation distances and vary the width aperture of one slit. 
This procedure enables us to obtain the Gouy phase difference and the logarithmic negativity as a function of the slit aperture. 
We draw our concluding remarks in section IV.

\section{Biphotons double-slit interferometry}
\label{SectionII}

Consider a biphoton produced in a typical Type-I SPDC process undergoing diffraction through a double-slit. 
It can be described by an effective wavefunction \cite{howell, bp, ford} in the position space whose parameters encode phase space entanglement, and a propagator can be derived \citep{fourier, bp} to evolve the double-Gaussian biphoton wavefunction in time.
Analytical expressions for the time evolution and intensity at the detection screen can be obtained by considering Gaussian slit apertures \cite{cris}. 
From the four wavefunctions describing the biphoton diffraction through a double-slit (namely the two photons passing either through the same or different slits), we extract the Gouy phase, a geometric phase that arises from transverse spatial confinement.
Thereafter, we can calculate the logarithmic negativity using the photon pair diffracted state in terms of phase space correlations expressed by the covariance matrix.
Quantum correlations of a biphoton pair were firstly investigated in connection with the Gouy phase in \cite{Kawase}.
Later on, a relation between the Gouy phase and quantum correlations  through the logarithmic negativity for the biphoton free evolution was shown in \cite{crisfree}. 
Here, we express the relation between the Gouy phase and the logarithmic negativity through the double-slit geometrical parameters,  as seen in figure \ref{gouy_N_beta}. 
We also study how the position cross-correlations affect the interference pattern.
By judiciously choosing a regime for position correlations, we derive the visibility and relative intensity  \cite{bramon, Paz4}, which are experimentally accessible and contain information about the Gouy phase difference.

\subsection{Type-I SPDC photon pair state in a double-slit and the Gouy phase}

\begin{figure}[htp]
\centering
\includegraphics[height=4cm,width=5 cm]{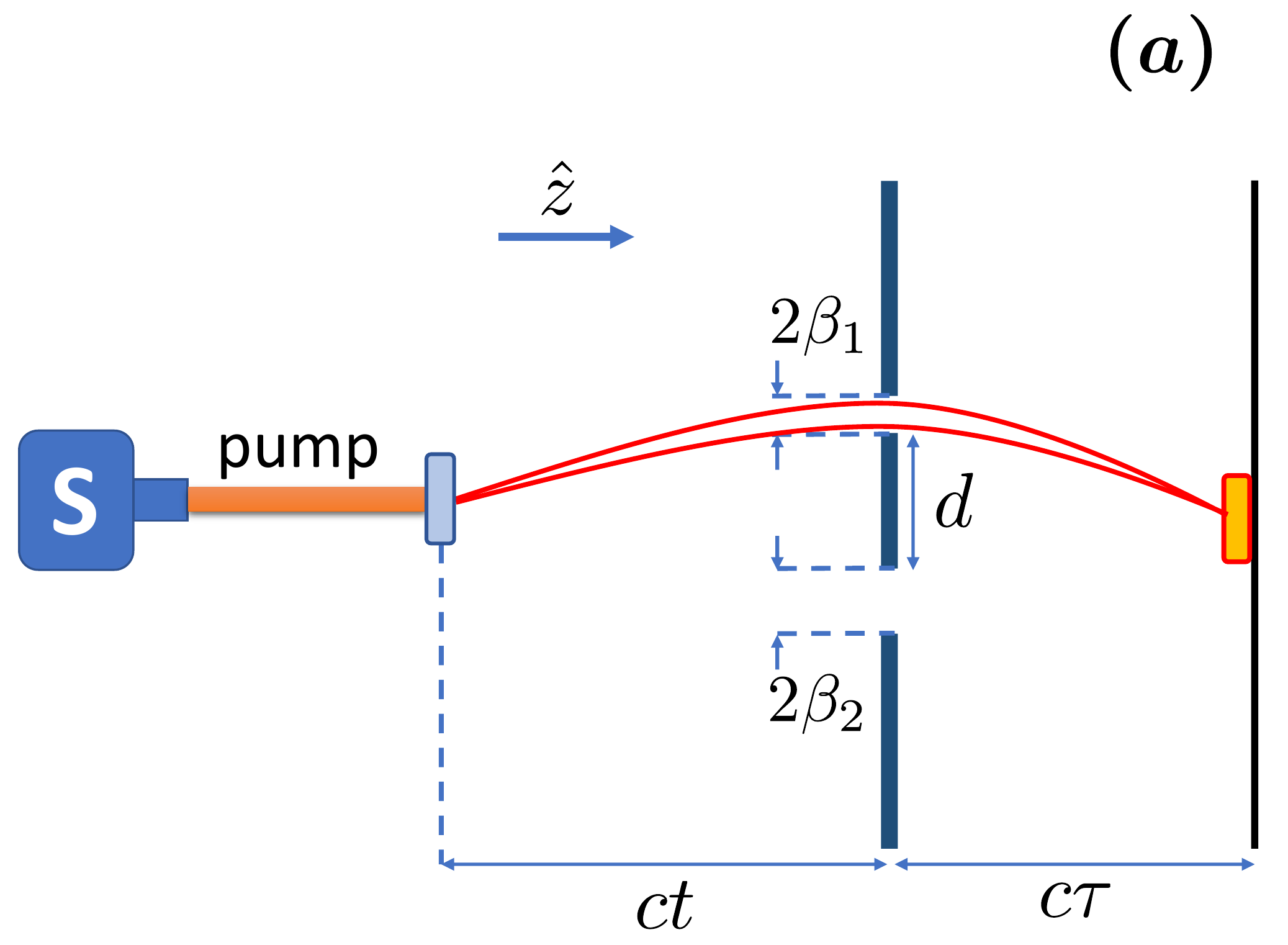}
\includegraphics[height=4cm,width=5 cm]{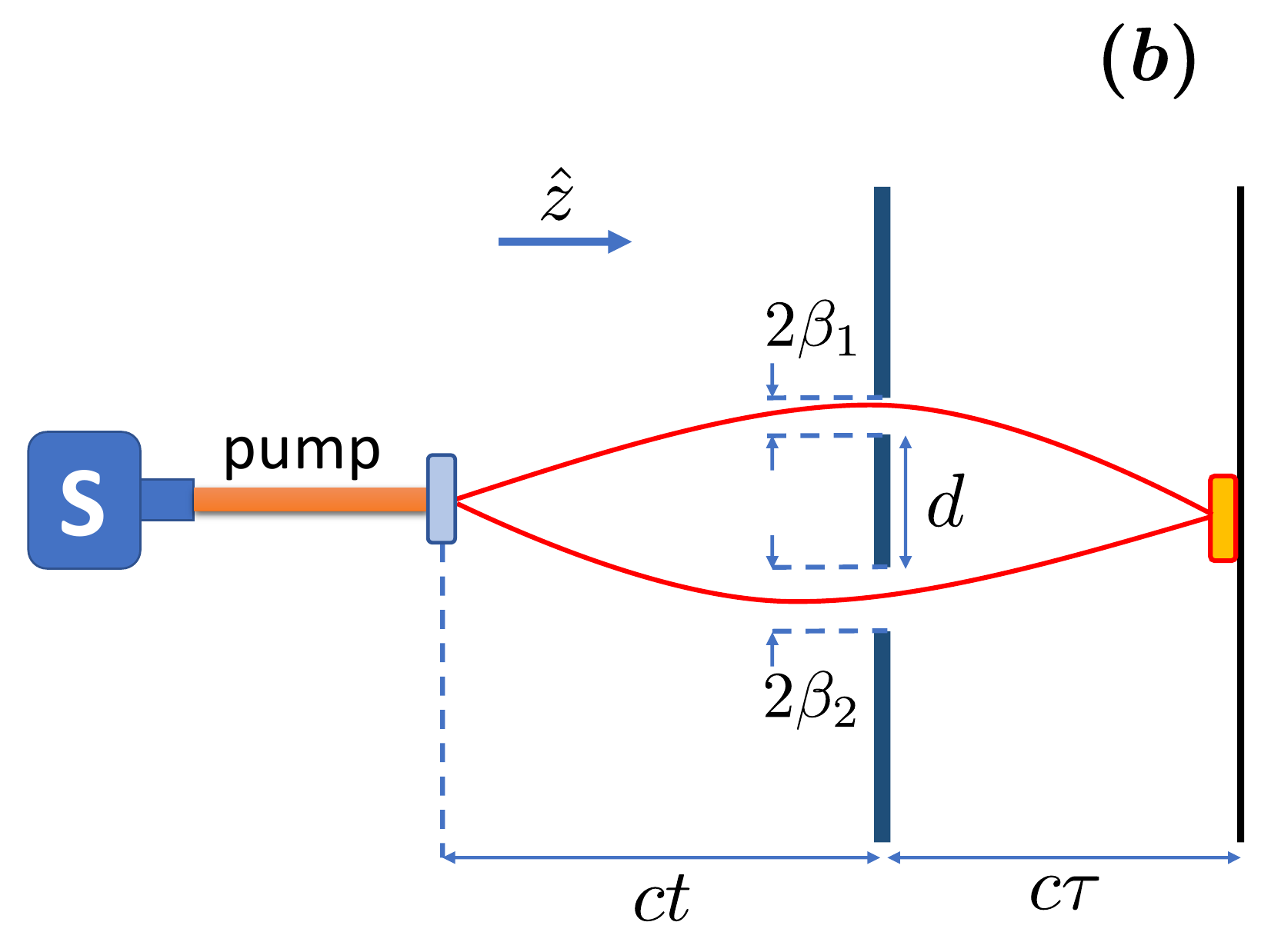}
\caption{Double slit setup for biphoton interference. (a) both photons can propagate through slit 1 (upper slit) or slit 2 (lower slit); (b) each photon of the pair can propagate through different slits. The photons travel a distance $z=ct$ from the nonlinear crystal to the slits of widths $\beta_1$ (upper slit) and $\beta_2$ (lower slit), with $d$ being the inter-slit distance. Thereafter, the photons travel a distance $z_{\tau}=c \tau$ until the detection screen.} \label{aparato}
\end{figure}

Let us firstly consider the free evolution of the biphoton wavefunction. 
The effective wavefunction for biphotons  generated in a degenerate collinear Type-I SPDC process is given by \cite{howell, bp, ford}
\begin{equation}
\psi_0(x_1, x_2)=\frac{1}{\sqrt{\pi \sigma \Omega}}
e^{\frac{-(x_1-x_2)^2}{4\sigma^2}}
e^{\frac{-(x_1+x_2)^2}{4\Omega^2}}, \label{estadoinicial}
\end{equation}
which is the generalized EPR state for the momentum-entangled particles.
In equation (\ref{estadoinicial}), $x_{1,2}$ is one of the transverse biphoton coordinates for which the effective wavefunction factorizes in the paraxial approximation. 
Moreover, $\Omega$ and $\hbar/\sigma$ quantify the position and momentum spread of the particles in the $x$ direction \cite{bp}. 
Notice that this state is not entangled (it factorizes) if $\Omega=\sigma$ \cite{crisfree}. 
The time propagation kernel for each of the photons in the pair is given by
\begin{equation}
K_y (y , t ; y', t') = 
\sqrt{\frac{1}{i\lambda
ct}}\exp{\bigg[-\frac{2\pi(y-y')^2}{i\lambda c(t-t')}\bigg]},
\label{propagator}
\end{equation}
from which we write the state describing the free propagation of the biphoton as
\begin{equation}
 \psi (r,q,t) =\int_{r', q'} K_r(r,t;r',0) K_q(q,t;q',0) \psi_0(r',q'),
 \label{estadoinicial1}
 \end{equation}
where  $\psi(r',q')$ is given by equation (\ref{estadoinicial}) written in terms of relative coordinates $r'=(x_1+x_2)/2$ and $q'=(x_1-x_2)/2$. 
Writing the time propagation as a function of the longitudinal distance \cite{crisfree}, namely $z=ct$, yields
\begin{equation}\label{psifp}
 \begin{split}
\psi(r,q,z)=& \frac{1}{\sqrt{4\pi w(z)\tilde{w}(z)}}\exp \bigg
\lbrace -\bigg[\frac{ r^2}{w^2(z)} +\frac{q^2}{\tilde{w}^2(z)}
\bigg] \bigg \rbrace \\ & \times \exp \bigg \lbrace -i
\bigg[-\frac{k_0}{r_{+}}r^2- \frac{k_0}{r_{-}}q^2+\zeta(z)\bigg]
\bigg \rbrace.
\end{split}
\end{equation}
Similarly to the single particle propagation, the biphoton wavefunction is characterized by the wavepacket spreads $w(z)$ and
$\tilde{w}(z)$, the radius
of curvature of the wave fronts $r_{\pm}(z)$, and the biphoton Gouy phase for the free propagation $\zeta(z)$, whose expressions are presented in Appendix \ref{freeprop}.
In the double slit setup, the biphoton covers a distance $z=ct$ from the source to the slits and $z_{\tau}=c\tau$ from the slits to the detection screen. 
There are four contributions that account for the intensity at the detection screen: both entangled photons propagating through the same slit and each one propagating through different slits, as seen in figure \ref{aparato}.
The complete evolution from source to detection screen for a type-I SPDC biphoton is described by the wavefunction:

\begin{equation}
\begin{split}
\label{integral}
 \Psi(r,q,z,z_{\tau}) = \int_{r',q'} & K_r(r, z+z_{\tau}; r', z) K_q(q,z+z_{\tau}; q', z) \nonumber \\ &\times
 F(r', q')  \psi(r',q',z),
 \end{split}
\end{equation}
where the integrals over the primed variables \{$r'$, $q'$\} run from $-\infty$ to $+\infty$. The propagators $K_r$ and $K_q$, and the initial state $\psi (r,q,z)$ are given by equations (\ref{propagator}) and (\ref{psifp}), respectively. The Gaussian window functions, representing slits of widths  $\beta_1$ (upper slit), $\beta_2$ (lower slit) and inter-slit center-to-center
distance $d$, read either
\begin{equation}\label{frqj}
F(r,q) \equiv\exp \left[-\frac{((r+q) \mp \frac{d}{2} )^2}{2\beta_{1(2)}^{2}}\right]
\exp\left[-\frac{((r-q) \mp \frac{d}{2} )^2}{2\beta_{1(2)}^2}\right],
\end{equation}
when the two photons travel through the same slit, where the upper (lower) sign refers to photons traveling by slit 1 (slit 2), or
\begin{equation}
F(r,q)  \equiv \exp \left[-\frac{((r+q) \mp \frac{d}{2} )^2}{2\beta_{1(2)}^2}\right]
\exp\left[-\frac{((r-q) \pm \frac{d}{2} )^2}{2\beta_{2(1)}^2}\right],
\end{equation}
when the two photons propagate through
different slits. 

Firstly, let us  consider identical slit apertures $\beta_{1}=\beta_{2}=\beta$. The wavefunction for the  biphoton propagating through the upper slit is
\begin{equation}\label{psiuu}
\begin{split}
\Psi_{uu} (r, q)&=\frac{1}{\sqrt{\pi B
\tilde{B}}}\exp\left[-\frac{(r-D_{uu}/2)^2}{B^2}\right]\exp\left[\frac{q^2}{\tilde{B}^2}\right]\\
& \times \exp\left(\frac{ik_0 }{ R_{+}}r^2+\frac{ik_0 }{
R_{-}}q^2+i \Delta_{uu} r + i\theta_{uu}+i\zeta\right),
\end{split}
\end{equation}
where the expressions for  $B$, $\tilde{B}$, $R_{\pm}$, $D_{uu}$, $\Delta_{uu}$ and $\theta_{uu}$ are written in Appendix \ref{appds}.
As for the biphoton propagating through the lower slit, $\psi_{dd}(r,q,z,z_{\tau})$, we replace $d$ with $-d$ in equation (\ref{psiuu}). On the other hand, the wavefunction describing a photon, say 1, passing through the upper slit and a photon 2 passing through the lower slit is
\begin{equation}\label{psiud}
\begin{split}
\Psi_{ud} (r, q)&=\frac{1}{\sqrt{\pi B \tilde{B}}}\exp\left[-\frac{r^2}{B^2}\right]\exp\left[-\frac{(q-D_{ud}/2)^2}{\tilde{B}^2}\right]\\
& \times \exp\left(\frac{ik_0 }{ R_{+}}r^2+\frac{ik_0 }{
R_{-}}q^2+i \Delta_{ud}q +i\theta_{ud}+i\zeta\right).
\end{split}
\end{equation}
Clearly the state $\psi_{du}(r,q,z,z_{\tau})$, which describes the propagation of photon 1
through the lower slit and photon 2 through the upper slit, is obtained by replacing $d$ with $-d$ in $\psi_{ud}(r,q,z,z_{\tau})$ (see Appendix \ref{appds}). Just as the biphoton propagation in the free space, $B(z,z_{\tau})$ and $\tilde{B}(z,z_{\tau})$ are the wavepacket spreads, $R_{\pm}(z,z_{\tau})$ is the radius of
curvature of the wave fronts for the propagation through the slit,
$D(z,z_{\tau})$ is the wavepacket separation and $\theta(z,z_{\tau})$ is a phase which depends on the propagation
distances and $d$. Therefore, the Gouy phase is given by
\begin{equation}\label{gouy_slit}
\begin{split}
\zeta=& -\frac{1}{2}\arctan\bigg[\frac{f(z,z_\tau, \beta) +g(z,z_\tau, \beta)}{1-f(z,z_\tau, \beta)g(z,z_\tau, \beta)}\bigg],
\end{split}
\end{equation}
where the functions $f(z,z_\tau, \beta)$ and $g(z,z_\tau, \beta)$ are given in Appendix \ref{appds}.
As the parameters $\Omega$ and $\sigma$ become identical, the initial state is not entangled \cite{bp} yet some correlations remain, as seen from the covariance matrix \cite{crisfree}.

\begin{figure}[ht!]
 \centering
\includegraphics[height=4.5cm,width=5.5cm]{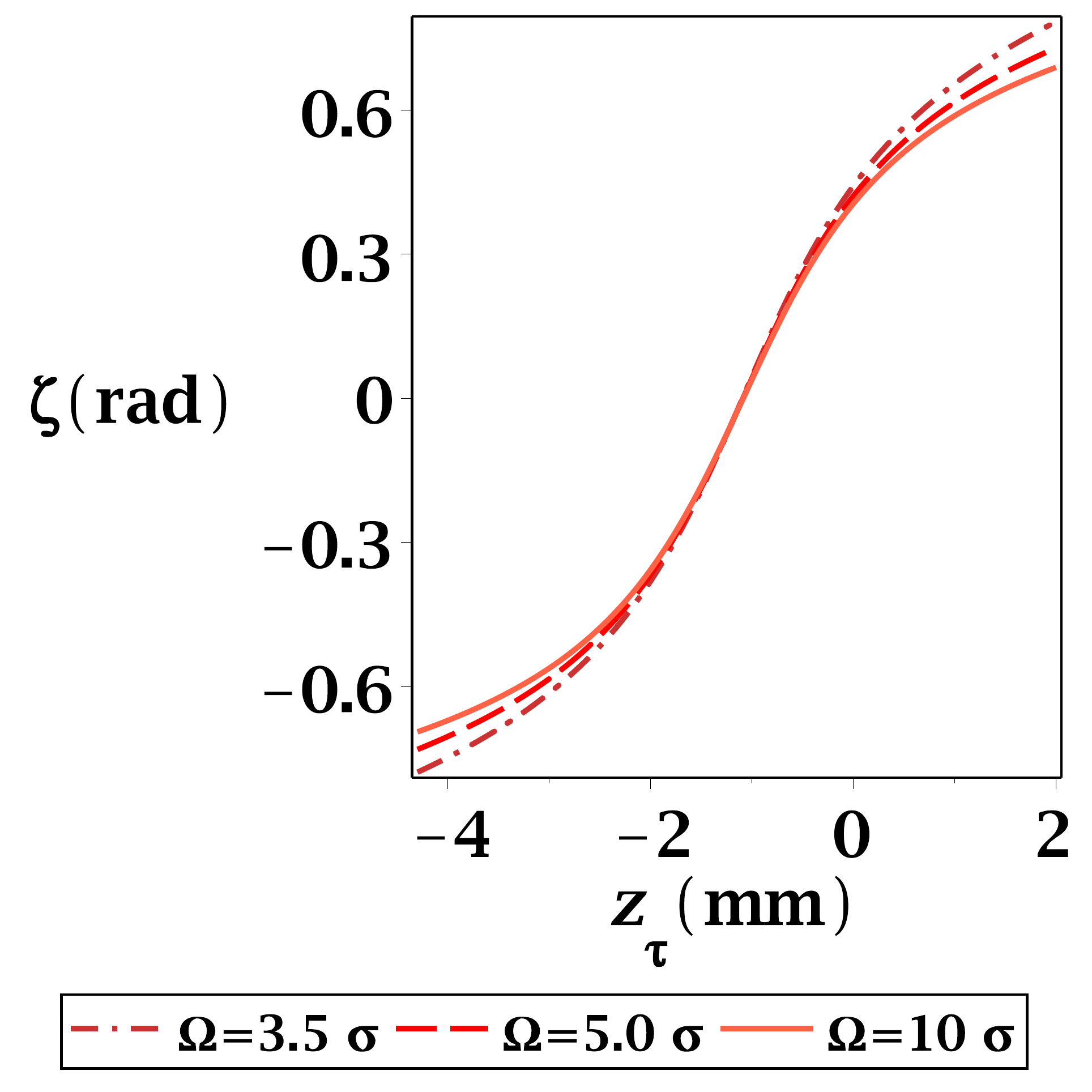}
\includegraphics[height=4.5cm,width=5.5cm]{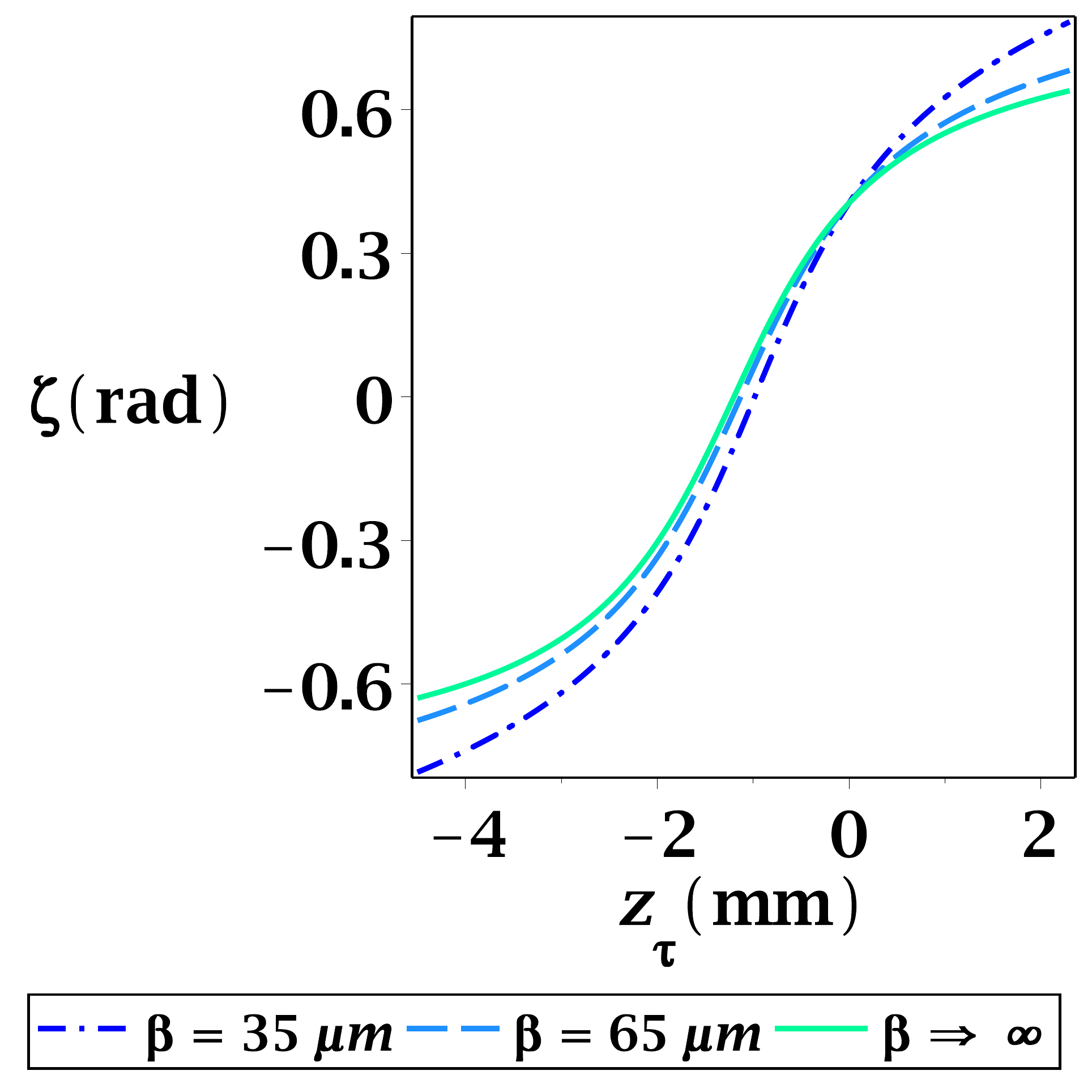}
\caption{Gouy phase as a function of the propagation distance from the slits to the detector $z_{\tau}$.
As the biphoton crosses the slits, its wavefunction acquires a Gouy phase that is dependent on the initial correlations. Notice that the  smaller the initial correlation, the larger is the total Gouy phase variation.
(Top) The  dashed curve corresponds to the Gouy phase for the parameter $\Omega=3.5\sigma$ whereas the dash-dotted curve  corresponds to $\Omega=5\sigma$. After diffraction, the Gouy phase shows dependence on the slit parameters.  
(Bottom) The Gouy phase for slit widths $\beta=35$ $\mu$m and $\beta=65\;\mathrm{\mu m}$ represented by the dash-dotted and dashed curves, respectively. Notice that as the slit width $\beta$ increases, the curve approaches the one for the  Gouy phase for free biphoton propagation ($\beta \rightarrow \infty$) represented by the solid curve.}
\label{gouy_tau}
\end{figure}

Let us study the behavior of the biphoton Gouy phase in equation (\ref{gouy_slit}) as a function  of the distance after slits $z_{\tau}$, for different values of the initial correlation between the two photons. 
A small correlation $\Omega=3.5\sigma$ produces a large total Gouy phase variation as compared with $\Omega=5.0\sigma$ or $\Omega=10\sigma$, represented by the dashed, dash-dotted and solid curves respectively, in the upper plot of figure \ref{gouy_tau}. 
This result has a similar pattern to the biphoton free propagation case \cite{crisfree},  showing that initial correlation between the photons plays a relevant role in the value of the Gouy phase  in equation (\ref{gouy_slit}), even when the wavepacket suffers diffraction.
Moreover, we analyse the Gouy phase (\ref{gouy_slit}) as a function of $z_{\tau}$, for three values of the slit width $\beta$ as seen in figure \ref{gouy_tau}. We observe that the smaller slit width, the larger the total Gouy phase variation (dash-dotted curve) as seen in figure \ref{gouy_tau}. In the limit where $\beta\rightarrow \infty$, the Gouy
phase in (\ref{gouy_slit}) for the propagation through a double-slit,
tends to the Gouy phase given by equation (\ref{gouy_free}) for the free propagation, as it should. 
We considered the following set values of parameters to construct figure \ref{gouy_tau}: 
biphoton wavelength $\lambda=702\;\mathrm{nm}$, laser pump wavelength $\lambda_p=351.1\;\mathrm{nm}$ and crystal typical length $L_z=7.0\;\mathrm{mm}$. 
This enables us to obtain $\sigma=\sqrt{\frac{L_p \lambda_p}{6 \pi}}=11.4\;\mathrm{\mu m}$ and
$z_{0-}=k_0\sigma^{2}=1.4\;\mathrm{mm}$, where $k_0=2\pi/ \lambda$.
We consider the distance from the source to the
slits $z=1.2\;\mathrm{mm}$. 
The slit width for the upper plot in figure \ref{gouy_tau} is $\beta=40\;\mathrm{\mu m}$.
The Gaussian  spread is such that $\Omega=10\sigma$ in the lower plot of figure \ref{gouy_tau}. 
Those are typical values as used in the biphoton double-slit experiment in \cite{brida}. 
It is noteworthy that  SPDC biphotons were  theoretical and experimentally studied by Kawase and collaborators in \cite{Kawase}, and the Gouy phase generated in the free evolution was connected to quantum correlations encoded in Laguerre-Gaussian modes. 
Moreover, the authors suggest that the Gouy phase can be used as a tool to manipulate multidimensional photonic quantum states. By diffracting a wavepacket through a double slit, the Gouy phase shift can be measured only in asymmetric setup, namely for different slit widths.
In the next section, we study this configuration and how the Gouy phase difference is related to the relative intensity and visibility. 

\subsection{Logarithmic negativity for type-I SPDC biphotons in a double-slit}
\label{gouyN}
A type-I SPDC biphoton pair diffracted through a double-slit is effectively described by the double-Gaussian state expressed by equation (\ref{psiuu}). Such Gaussian states are fully characterized by their first and second moments. It is possible to set the first moments to zero by local unitary operations, whilst keeping the entanglement unchanged, and the second moments are given by the covariance matrix elements \cite{Vidal}. 
A necessary condition for an entanglement quantifier is that it has to vanish if the state is separable.
According to the Peres-Horodecki criterion \cite{PHC}, if a state is separable the partial transpose of the correlation (variance) matrix of the state has a non-negative spectrum. Therefore, we can establish that a Gaussian state is separable if and only if the minimum value of the symplectic spectrum of $M^{T_2}$, $M$ defined in equation (\ref{covariance}), is greater than $1/2$, the lowest value allowed by the uncertainty principle \cite{ford}. 
Consider the wavefunction for the biphoton
propagating through the upper slit, as seen in figure \ref{aparato}(a). Using equation (\ref{psiuu}), we calculate the elements of the covariance matrix whose symplectic form can
be written as
\begin{eqnarray}\label{covariance}
M = \left[
\begin{array}{cccc}
g & 0 & c & 0 \\
0 & g & 0 & c' \\
c & 0 & h & 0 \\
0 & c' & 0 & h
\end{array}
\right]
\end{eqnarray}
which is related to the phase space correlation matrices
\begin{equation}
\begin{split}
G=& \begin{bmatrix}
\frac{\langle x_1^2\rangle}{L^2}  & \frac{\langle x_1p_1 +p_1x_1 \rangle}{2 \hbar}  \\
\frac{\langle x_1p_1 +p_1x_1 \rangle}{2 \hbar} & \frac{L^2\langle
p_1^2 \rangle}{\hbar^2} \nonumber
\end{bmatrix}, \\
H=&
\begin{bmatrix}
 \frac{\langle x_2^2\rangle}{L^2}  & \frac{\langle x_2p_2 +p_2x_2 \rangle}{2 \hbar}  \\
\frac{\langle x_2p_2 +p_2x_2 \rangle}{2 \hbar} & \frac{L^2\langle
p_2^2 \rangle}{\hbar^2} \nonumber
\end{bmatrix},\\
 &C=
\begin{bmatrix}
 \frac{\langle x_1 x_2 \rangle}{L^2}  & \frac{\langle x_1p_2  \rangle}{ \hbar}  \\
\frac{\langle x_2p_1 \rangle}{\hbar} & \frac{L^2\langle p_1 p_2
\rangle}{\hbar^2}
\end{bmatrix},
\end{split}
\end{equation}

\vspace{0.3cm} \normalsize \noindent through simple relations $\det
G= g^2$, $\det H= h^2$ and $\det C= cc'$. The constants $\hbar$ and
$L$, which appear in the above matrices, are inserted to make the
matrix $M$ dimensionless.  The full expressions for the elements of $M$ are disposed in Appendix \ref{covmat}. With that in sight, the logarithmic negativity is a suitable measure to quantify entanglement for general  Gaussian states. It can be written as $E_N=\text{max} \lbrace0,-\log(2 \nu_{min})\rbrace$ where,  $\nu_{min}$ is the lowest symplectic eigenvalue of
$M^{T_2}$. 
The equation that determines the symplectic eigenvalues is
$\nu^4+(g^2+c^2-2 cc')\nu^2+ \text{det} (M)=0$, with solutions $\pm
i\nu_{\alpha}$, $\alpha=1,2$ and $\nu_{\alpha}$ is the symplectic spectrum.
Thus, the logarithmic negativity obtained is expressed as

\begin{equation}\label{negatividade0}
\begin{split}
E_N=\ln \left[ \frac{\sqrt{2}icB\tilde{B}R_+R_-}{ \sqrt{\sqrt{-A_1A_2}+A_3}}\right],
\end{split}
\end{equation}
where $i$ is the imaginary unit and $c$ is the speed of light constant.
The expressions for $B$, $\tilde{B}$, $R_+$ and $R_-$ are shown in Appendix \ref{appds}, and the expressions for $A_1$, $A_2$ and $A_3$ are displayed in Appendix \ref{covmat}.
In a free evolution, the logarithmic negativity for a similar biphoton  \cite{crisfree} is written in terms of the Gaussian wavepacket spreads as  $E_N=\log{\big(\frac{\Omega}{\sigma}\big)}$, i.e. in terms only of the initial correlation between the photons. 
After diffracting through a double slit, the 
the logarithmic negativity calculated for the biphoton wavefunction depends on the geometrical  double-lit parameters as well. 
Consequently, we may explore how the quantum correlations represented by the logarithmic negativity behave as a function of the apparatus parameters. 
\begin{figure}[htp]
\centering
\includegraphics[height=4cm,width=4.0 cm]{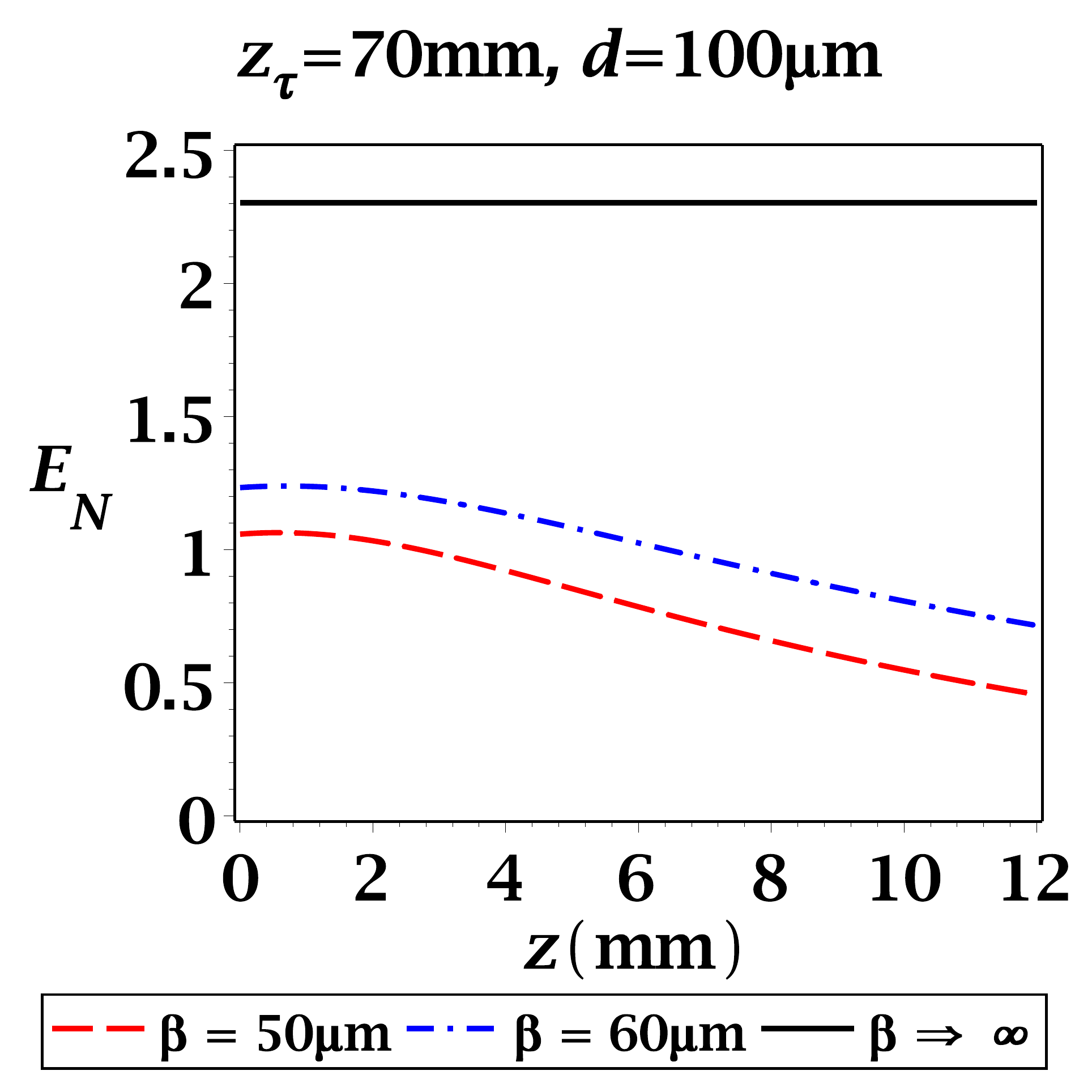}
\includegraphics[height=4cm,width=4.0 cm]{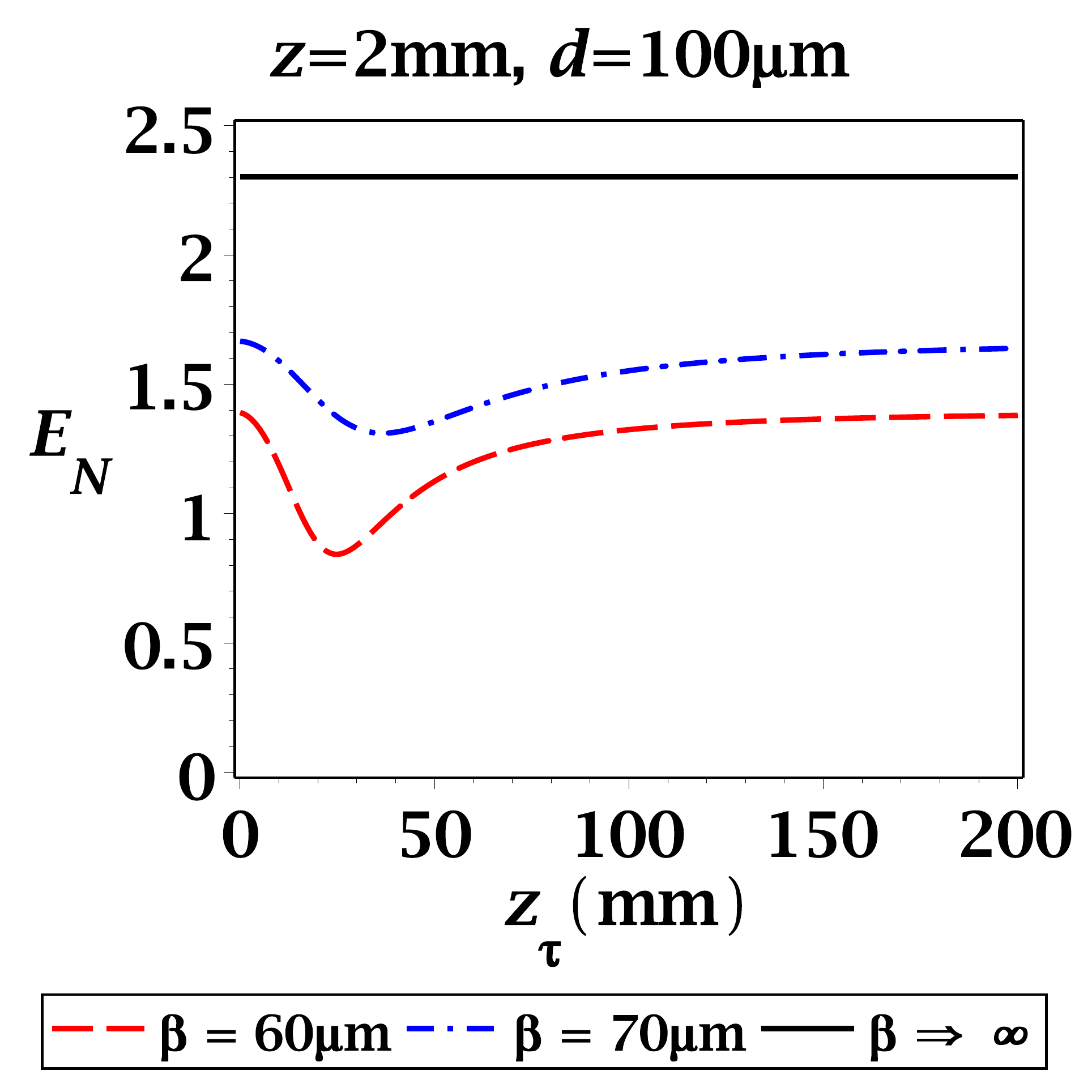}\\
\includegraphics[height=4cm,width=4.0 cm]{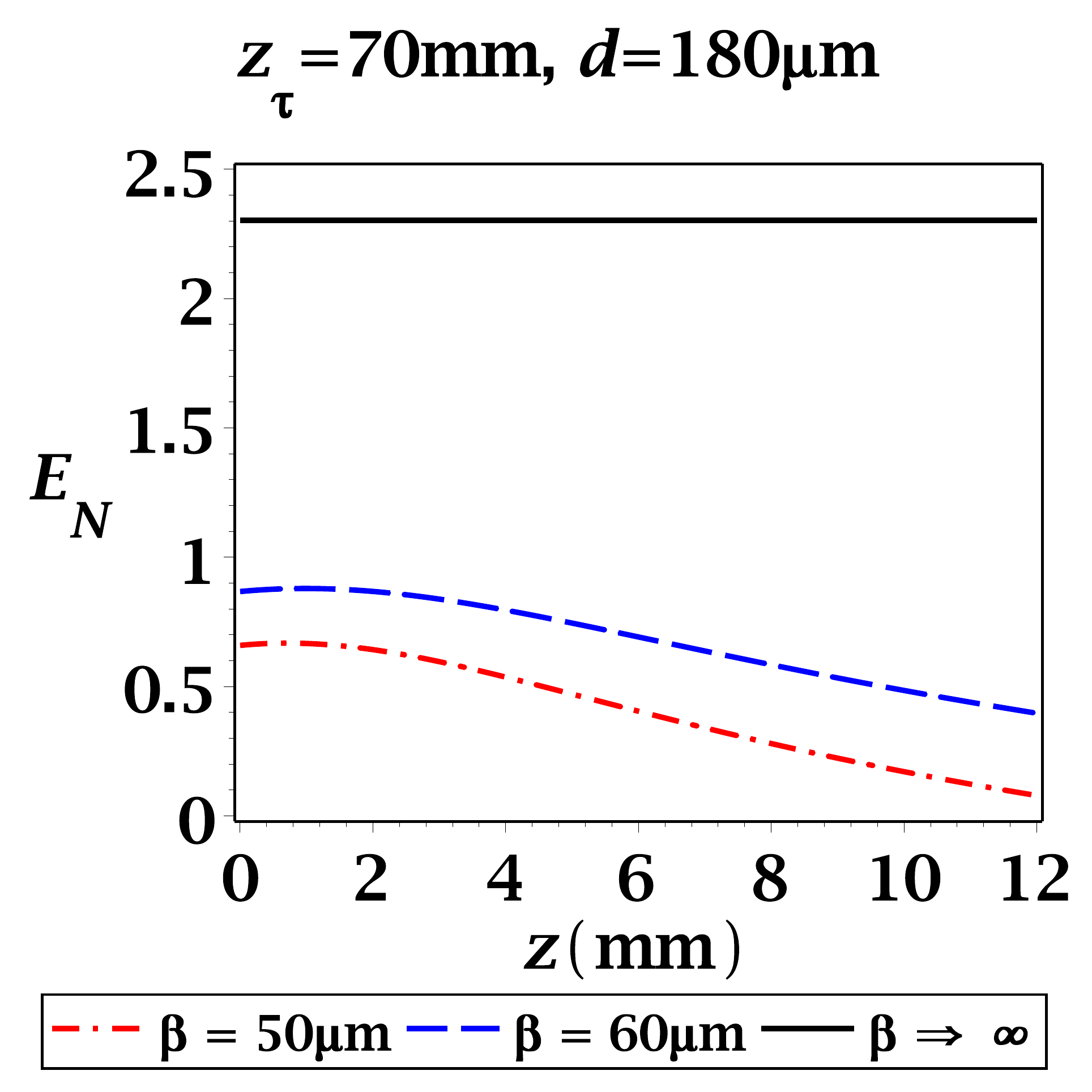}
\includegraphics[height=4cm,width=4.0 cm]{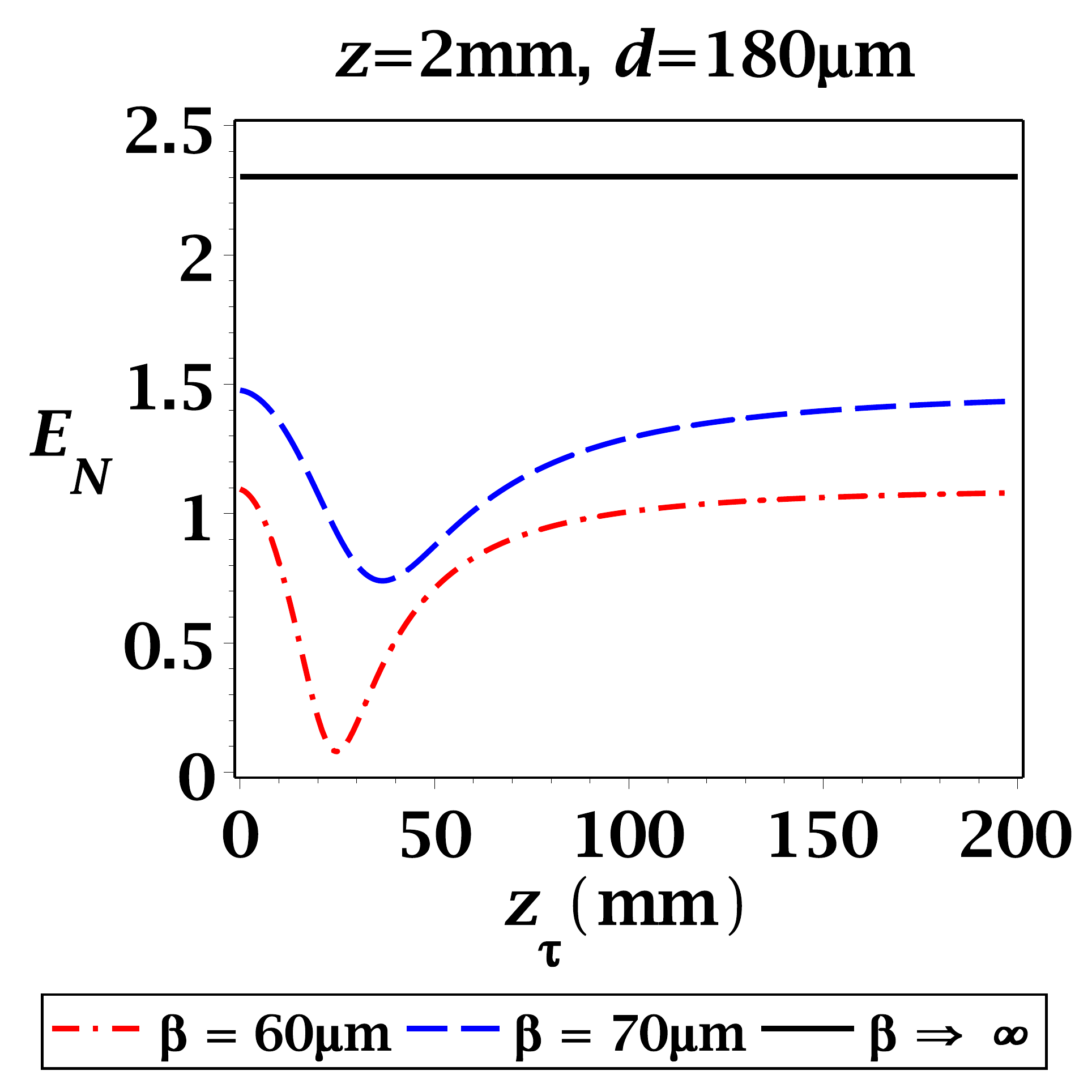}\\
\caption{The biphoton logarithmic negativity equation (\ref{negatividade0}) depends on propagation distances $z$ and $z_\tau$.
As the slit width increases, the quantum correlations become
independent of $z$ and $z_\tau$ and saturate at a maximum constant value, represented by the solid curve in the four plots. In these plots we have the logarithmic negativity $E_N$ as a function of the distance from the source to the slits $z$ (distance from the slits to the detector $z_{\tau}$), with $z_{\tau}=70\;\mathrm{mm}$ ($z=2\;\mathrm{mm}$), for the slit widths $\beta=50\;\mathrm{mm}$ and $\beta=60\;\mathrm{\mu m}$ ($\beta=60\;\mathrm{\mu m}$ and $\beta=70\;\mathrm{\mu m}$), dash-dotted and dashed curve, respectively.
For two different values of slit distance $d=100\;\mathrm{\mu} m$ (upper plots) and $d=180\;\mathrm{\mu} m$ (lower plots), we observe deeper valleys for larger values of $d$ due to loss of correlation for a range of values of $z_{\tau}$. }
 \label{N_z_ztau}
\end{figure}

In order to study how $E_N$ is affected by the slit parameters, we show in the graphs of figure \ref{N_z_ztau} the logarithmic negativity as a function of  $z$ and $z_{\tau}$, the flight distances before and after the slits, respectively, for some typical values of the inter-slit distance $d$ and the slit width $\beta$.  
Notice that the logarithmic negativity $E_N$ decreases as  $z$ increases (plots on the left) whereas, as $z_{\tau}$ increases, it displays a valley  before it becomes constant (plots on the right).
As the inter-slit distance $d$ increases, the photons become less correlated, and the logarithmic negativity is nearly zero for a slit width $\beta=60\;\mathrm{\mu m}$ (dash-dotted curve in the lower-right plot in  figure \ref{N_z_ztau}). 
  One can see this by comparing the upper-right with the lower-right plot, the only difference being that $d=100\;\mathrm{\mu m}$ in the upper one and  $d=180\;\mathrm{\mu m}$ in the lower one. 
  Due to the existence of a region where the momentum uncertainty has a maximum, between $z_{\tau}=0$ and  $z_{\tau}=50\;\mathrm{m m}$, the logarithmic negativity decreases in that same region, which is consistent with the  fact of decreased entanglement leads to higher uncertainty of the position and momentum \cite{berta}.

Moreover, notice in the plots of figure \ref{N_z_ztau} that as the slit width $\beta$ increases, so does $E_N$. Evidently when $\beta\rightarrow \infty$, $E_N$ becomes independent of $z$ and $z_\tau$ and saturates at a maximum constant value. This value coincides with logarithmic negativity for biphotons in a free propagation regime $E_N=\log_{e}{\big(\frac{\Omega}{\sigma}\big)\approx2.3}$. We used a biphoton wavelength  $\lambda=702\;\mathrm{nm}$, laser pump wavelength $\lambda_p=351.1\;\mathrm{nm}$ and a  crystal typical length $L_z=7.0\;\mathrm{mm}$. Thus, we have $\sigma=\sqrt{\frac{L_p \lambda_p}{6 \pi}}=11.4\;\mathrm{\mu m}$, $\Omega=10\sigma$   and
$z_{0-}=k_0 \sigma^{2}=1.4\;\mathrm{mm}$, where $k_0=2\pi/ \lambda$
\cite{brida}. 

\begin{figure}[htp]
\centering
\includegraphics[height=4cm,width=4.0 cm]{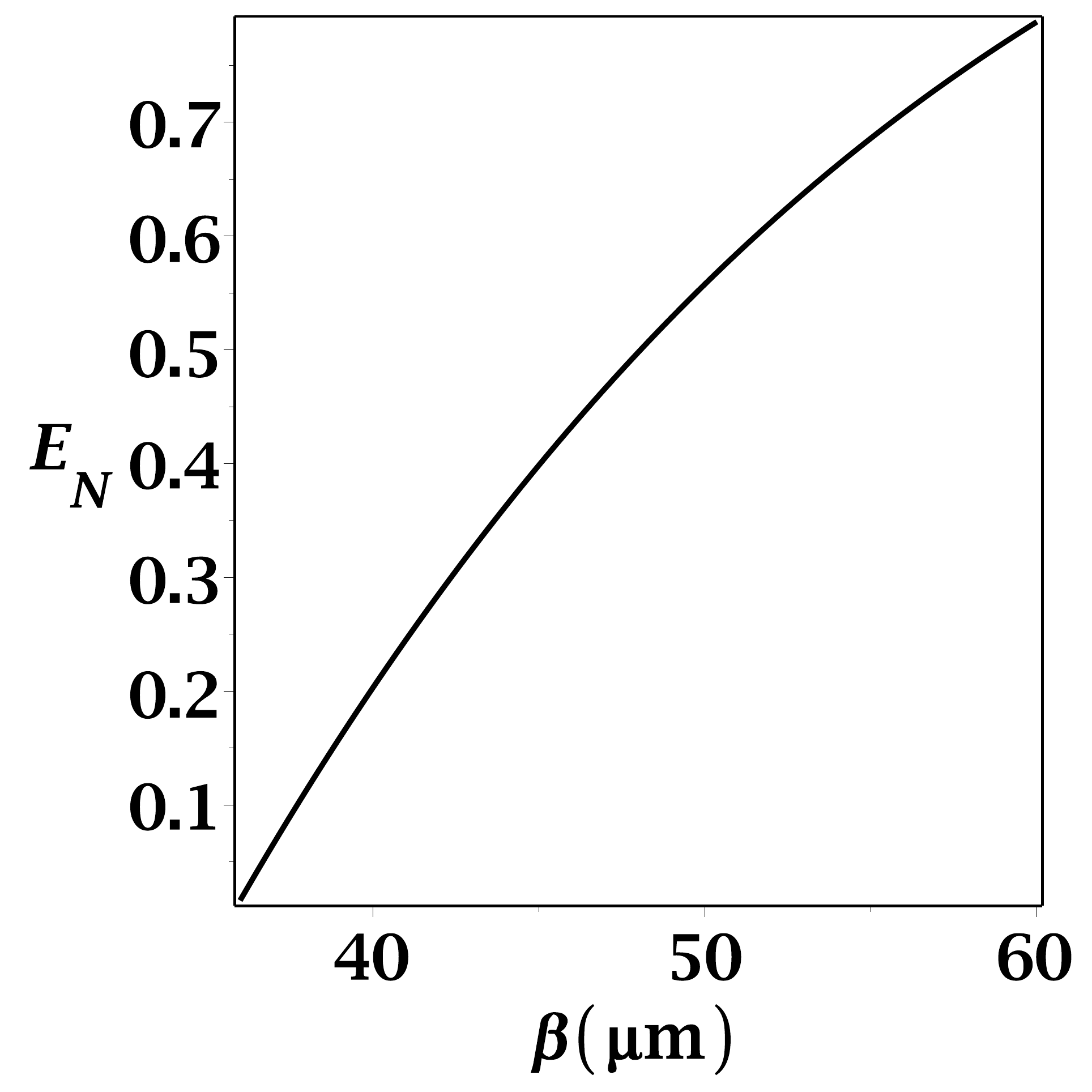}
\includegraphics[height=4cm,width=4.0 cm]{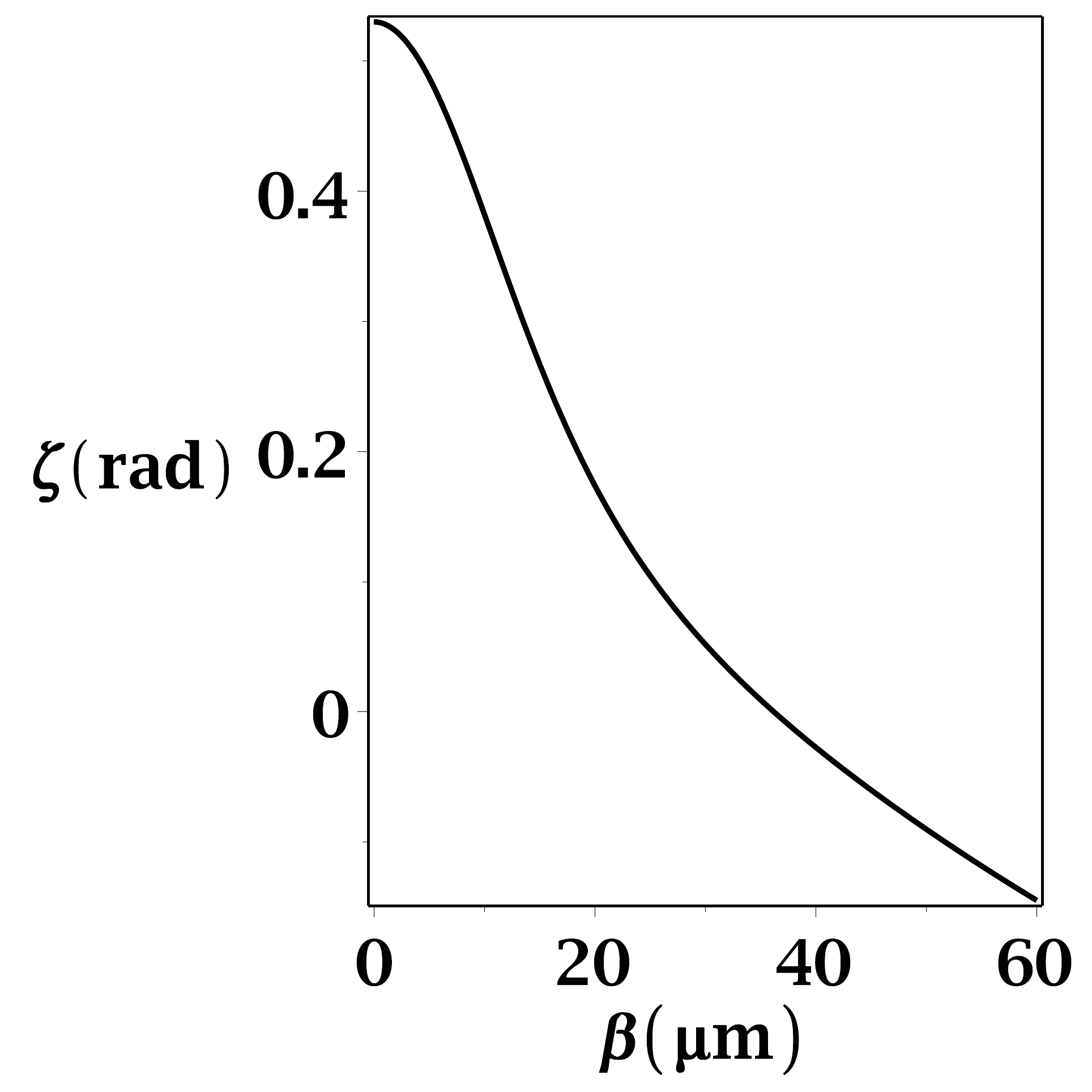}
\includegraphics[height=4cm,width=4.0 cm]{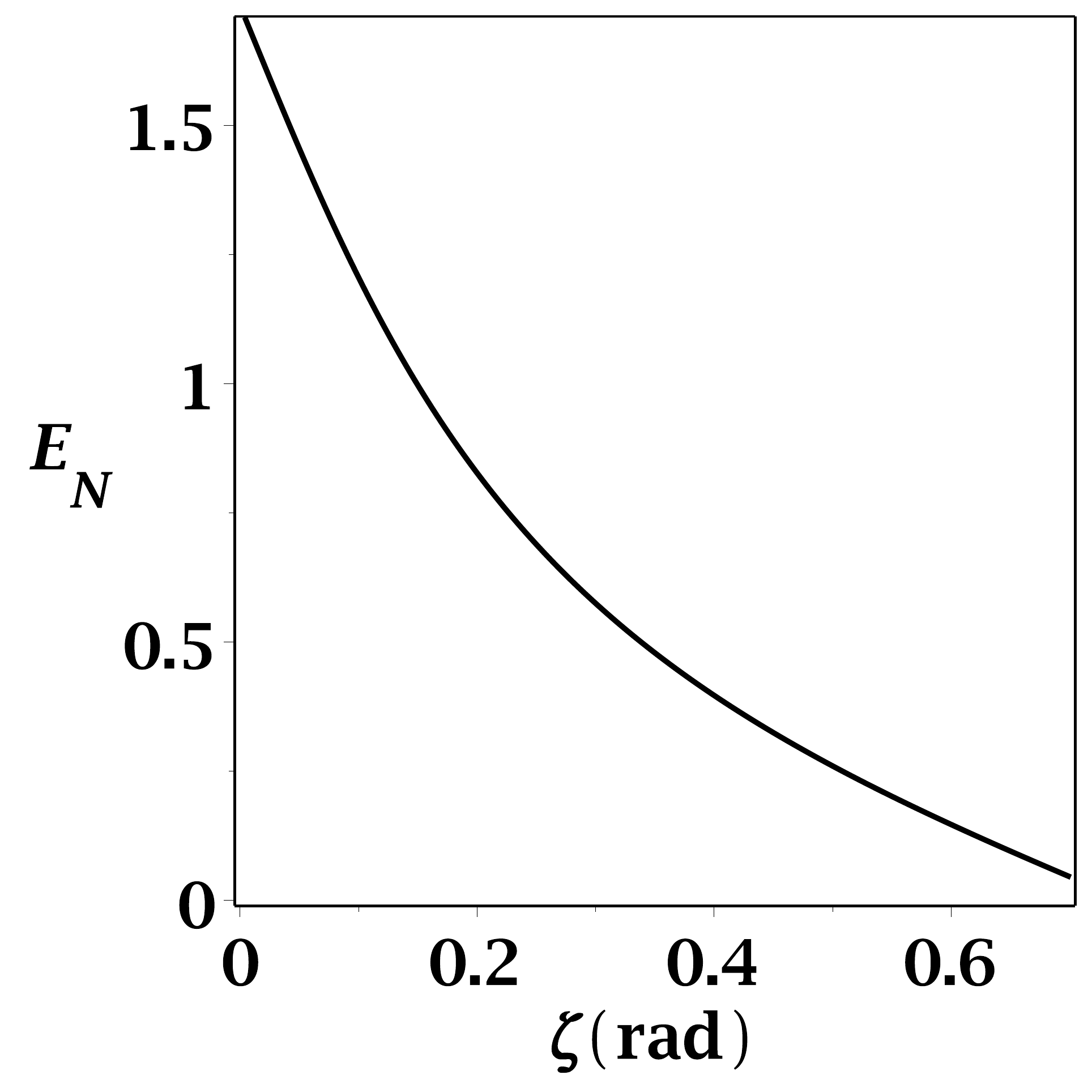}
\caption{The biphoton entanglement at the detection screen is influenced by the transverse spatial confinement represented by the slit. As we can see, the logarithmic negativity $E_N$ equation (\ref{negatividade0}), for $d=0$, as a function of the slit width $\beta_{1}=\beta_{2}=\beta$ indicates that the larger is $\beta$, the larger is the biphoton entanglement at the detection screen (upper-left plot). On the other hand, the biphoton Gouy phase variation $\zeta$ increases with transverse spatial confinement (upper-right plot). Finally, we verify that the entanglement represented by $E_N$ decreases as the Gouy phase $\zeta$ increases (lower plot).}
 \label{gouy_N_beta}
\end{figure}

While the logarithmic negativity depends on all geometrical parameters of the double-slit, the Gouy phase does not  depend on the inter-slit distance $d$. Later, we show that the Gouy phase difference, in terms of the relative intensity and visibility, does present a dependence on $d$. Then, to investigate the relation between the biphoton Gouy phase and its quantum correlations, we may replace for a while the two slits with a single one, located in the origin $x=0$, by setting  $d=0$. On the other hand, the slit width $\beta$ plays an important role in the measurement of the Gouy phase, as expected. Consequently, the biphoton logarithmic negativity at the detection screen is influenced by the spatial transverse confinement represented by $\beta$. The logarithmic negativity $E_N$ as a
function of slit width $\beta$ is exhibited in figure \ref{gouy_N_beta} (upper-left plot). Notice that the smaller $\beta$, so are the quantum correlations between the photons. We have also depicted the Gouy phase in equation (\ref{gouy_slit}) as a function of  $\beta$ in figure \ref{gouy_N_beta} (upper-right plot), which shows that the Gouy phase diminishes as the slit width increases.
We used a biphoton wavelength $\lambda=702\;\mathrm{nm}$, laser pump wavelength $\lambda_p=351.1\;\mathrm{nm}$ and the crystal typical length
$L_z=7.0\;\mathrm{mm}$, in figure \ref{gouy_N_beta}, which yield $\sigma=\sqrt{\frac{L_p \lambda_p}{6 \pi}}=11.4\;\mathrm{\mu m}$ and
$z_{0-}=k_0\sigma^{2}=1.4\;\mathrm{mm}$, where $k_0=2\pi/ \lambda$
\cite{brida}. We also considered $\Omega=10\sigma$, $d=200\;\mathrm{mm}$,
$z=2\;\mathrm{mm}$ and $z_{\tau}=70\;\mathrm{mm}$.

We now proceed to investigate the relation between the phase space quantum correlations and the Gouy phase of a type-I SPDC biphoton in a double slit setup. For a free evolution, such a relation was first studied in  \cite{Kawase}, where the authors used lenses to focus the biphoton beams and thus manipulated the quantum correlations in the free propagation evolution by promoting transverse spatial confinement of the wavepacket. In our proposal, the Gouy phase difference arises due to the transverse spatial confinement promoted by a double slit, and it can be  measured if we use different slit widths $\beta$. Since in the case we are considering here both the logarithmic negativity and the Gouy phase depend on the slit parameters, we can relate them through the slit width (see lower plot in figure \ref{gouy_N_beta}). For the set of parameters considered in figure \ref{gouy_N_beta}, the logarithmic negativity decreases as a function of the Gouy phase. Therefore, we conclude that for the system type-I SPDC biphoton diffracted by a double slit, one can access information about the logarithmic negativity through values of the Gouy phase. In the next section, we explore the possibility of relating the logarithmic negativity and the Gouy phase difference in an asymmetric double slit experiment. For this purpose, we have to define the visibility and the relative intensity, which is attainable in a certain regime of position correlations. 

\subsection{Position cross-correlation effects in the interference pattern}

In this section we will see that by choosing the wavepacket parameters such that one has  strong position correlation at the slits, we can numerically disregard the amplitudes that correspond to photons crossing different slits. In turn, this configuration results in a slightly anti-correlated photon pair at the detection screen. Also, in analogy to a single particle double slit interferometric setup, we can define the visibility and the relative intensity \cite{bramon, Paz4}.
We investigate how the interference pattern of biphotons diffracted through a double slit is related to their position correlations at the detection screen. 

The two-particle normalized spatial cross-correlation reads \cite{rho}
\begin{equation}\label{rho}
\rho_x=\frac{\langle x_1x_2\rangle-\langle x_1\rangle \langle
x_2\rangle}{\sigma_{x1}\sigma_{x2}},
\end{equation}
where $\sigma_{x_{1,2}}$ is the standard deviation of $x_{1,2}$. It ranges from $-1$ to $1$, it is zero if the two particles are uncorrelated. They are spatially closely correlated (bunched), if $\rho_x \rightarrow 1$ and  spatially
closely anti-correlated, if $\rho_x \rightarrow -1$. The position cross-correlations at the detector, calculated using the wavefunctions  in equation (\ref{psiuu}) and  equation (\ref{psiud}), respectively, are
\begin{equation}\label{rhouudd}
\begin{split}
\rho_{uu}(z,z_\tau,\beta)=\frac{B-\tilde{B}+D_{uu}}{B+\tilde{B}+D_{uu}},\\ \\
\rho_{ud}(z,z_\tau,\beta)=\frac{B-\tilde{B}-D_{ud}}{B+\tilde{B}+D_{ud}}.
\end{split}
\end{equation}
As we can observe, these quantities are expressed in terms of the
wavepacket spreads $B$ and $\tilde{B}$, as well as the wavepacket separation
$D_{uu}$ and $D_{ud}$ (whose expressions are in Appendix \ref{appds}). Also in the limit of wide slit width $\beta\rightarrow \infty$, we recover the biphoton cross-correlation for a
free propagation 
\begin{equation}
\rho(z,z_\tau)=\frac{\left(\Omega^{2}-\sigma^{2}\right)}{\left(\Omega^{2}+\sigma^{2}\right)}\frac{\left[1-\left(\frac{ z+z_{\tau}}{k_0\sigma\Omega}\right)^{2}\right]}{\left[1+\left(\frac{ z+z_{\tau}}{k_0\sigma\Omega}\right)^{2}\right]},
\end{equation}
where, as usual, $\Omega$ and $\sigma$ are the wavepacket initial parameters, $k_0=2\pi/\lambda$, $z$ and $z_{\tau}$ are free propagation distances traveled by the photons.
\begin{figure}[ht!]
\centering
\includegraphics[width=5cm]{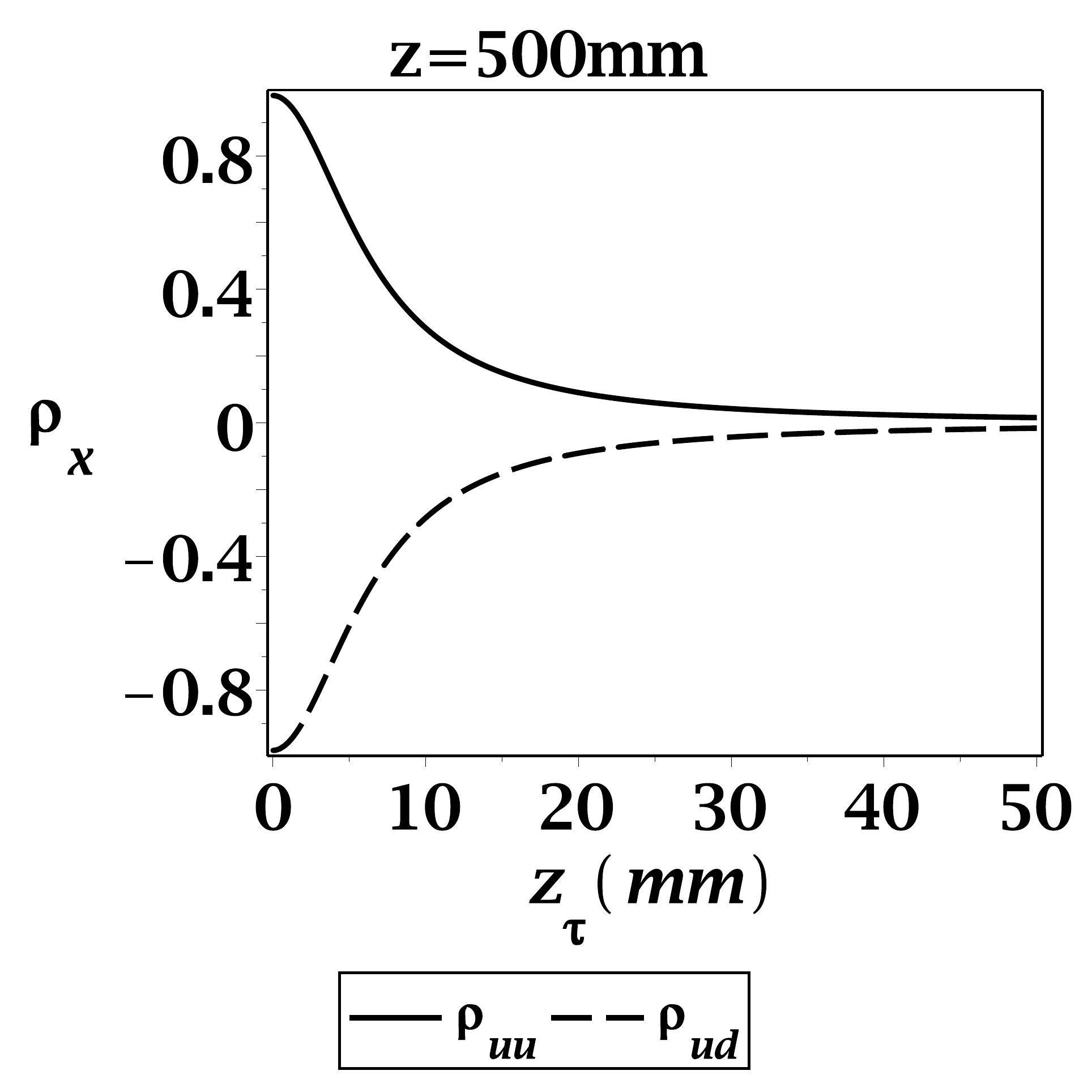}
\includegraphics[width=5cm]{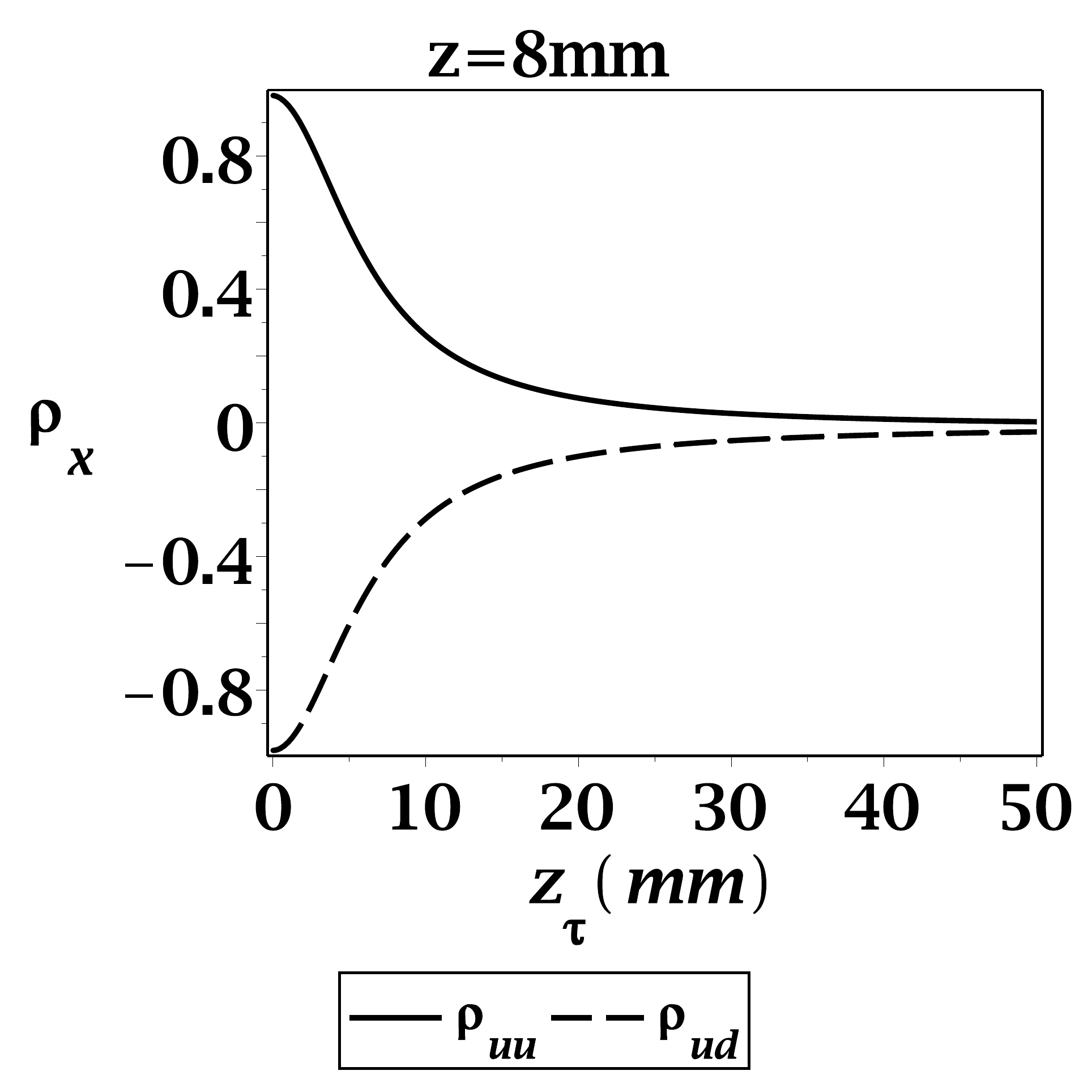}\\
\caption{
Cross-correlation $\rho_{uu}$ ($\rho_{ud}$) as a function of the distance from the slits to the detector $z_{\tau}$, calculated using the wavefunction for the two photons crossing the upper slit (crossing different slits). (Top) The cross-correlations for  $\rho_{uu}$  and $\rho_{ud}$ for the distance between the source and the slits $z=500\;\mathrm{mm}$.  (Bottom) The cross-correlations for  $\rho_{uu}$  and $\rho_{ud}$ for the distance between source and the slits $z=8\;\mathrm{mm}$
We can observe from both plots that the biphoton passing through the same slit is very correlated just after crossing the slits $z_{\tau}\approx0$ and become less correlated far from the slits. A closer view at $z_{\tau}\approx0$ shows that the pair becomes actually slightly anti-correlated. The constant values achieved by the cross-correlations are $\rho_{uu}=0.00035$ and $\rho_{ud}=0.000035$, for $z=500\;\mathrm{mm}$, and $\rho_{uu}=-0.0099$ and $\rho_{ud}=-0.010$, for $z=8\;\mathrm{mm}$.  }\label{correlation}
\end{figure}
For the position cross-correlations in (\ref{rhouudd}) and depicted in figure \ref{correlation}, we consider  $\sigma=11.4\;\mathrm{\mu m}$, $\Omega=10\sigma$, $d=100\;\mathrm{\mu m}$, $\beta=5.0\;\mathrm{\mu m}$.
Just after crossing the upper slit, the biphoton is evidently very correlated in its spatial coordinates. 
As the photons move away from the slits, the position correlation $\rho_{uu}$ decreases until reaching a constant value. 
In contrast, photons traveling  separately through different slits are considerably anti-correlated just after crossing. The position correlation $\rho_{ud}$ decreases as $z_{\tau}$ increases and the photons become less anti-correlated until $\rho_{ud}$ assumes a constant value. The asymptotic values of correlations can be easily obtained numerically. For instance if we take  $z_{\tau}=500\;\mathrm{mm}$,  we get $\rho_{uu}^{\infty}=0.00035$ and  $\rho_{ud}^{\infty}=0.000035$ for $z=500\;\mathrm{mm}$, whereas, for $z=8\;\mathrm{\mu m}$, we have $\rho_{uu}^{\infty}=-0.0099$ and $\rho_{ud}^{\infty}=-0.010$. 
In other words, the biphoton becomes uncorrelated as it travels large distances between the slits and the detector $z_{\tau}$, regardless of the distance traveled before reaching the slits $z$ (see figure \ref{correlation}). 
However, the asymptotic value reached by the cross-correlations is not equal to zero. Thus, we see that for $z=500\;\mathrm{\mu m}$, photons crossing the same slit are very correlated $\rho_{uu}=0.98$ whereas when each one crosses a different slit, they are as much anti-correlated, $\rho_{ud}=-0.98$,  just after crossing the slits. As $z_{\tau}$ increases, they loose spatial correlation in both cases. For small distances before the diffraction, say $z=8\;\mathrm{\mu m}$, the photons leave the slits with the position cross-correlations $\rho_{uu}=0.98$ and $\rho_{ud}=-0.98$, and they become anti-correlated at the detector far from the slits.

As we showed in subsection \ref{gouyN} (see figure \ref{gouy_N_beta}), one can access information about the logarithmic negativity through the values of the Gouy phase in this setup. In the next section, we obtain analytical expressions for an asymmetric double slit experiment in order to obtain (experimentally) measurable values of  Gouy phase differences of the biphoton.
For this purpose, we  need  analytical expressions for the visibility and the relative intensity. Such quantities are easily obtained analytically for one particle interference. Despite having a two particle wavefunction, we may still obtain analytical
expressions because within a particular range of position correlations only two wavefunctions instead of four contribute effectively to the interference pattern at the detection screen. In order to see how the position correlations affect the interference pattern at the detection screen, we study the intensity as a function of the position on the screen $x$, using different values of position correlation. The intensity composed by the four possibilities for the biphoton to cross the double-slit, the intensity including only the (two) wavefunctions that represent the photons passing through the same slit and the intensity for wavefunctions that represent only the photons passing through different slits are given, respectively, by
\begin{equation}
\begin{split}
I_{4\Psi}=&|\Psi_{uu}+\Psi_{ud}+\Psi_{du}+\Psi_{dd}|^2,\\ \\
\hspace{0.5cm} I_{2\Psi}=|\Psi_{uu}+\Psi_{dd}|^2&, \hspace{0.5cm} \text{and} \hspace{0.5cm}
I^{\prime}_{2\Psi}=|\Psi_{ud}+\Psi_{du}|^2.
\end{split}
\label{intensity}
\end{equation}

\begin{figure}[ht!]
 \centering
\includegraphics[width=4.2cm]{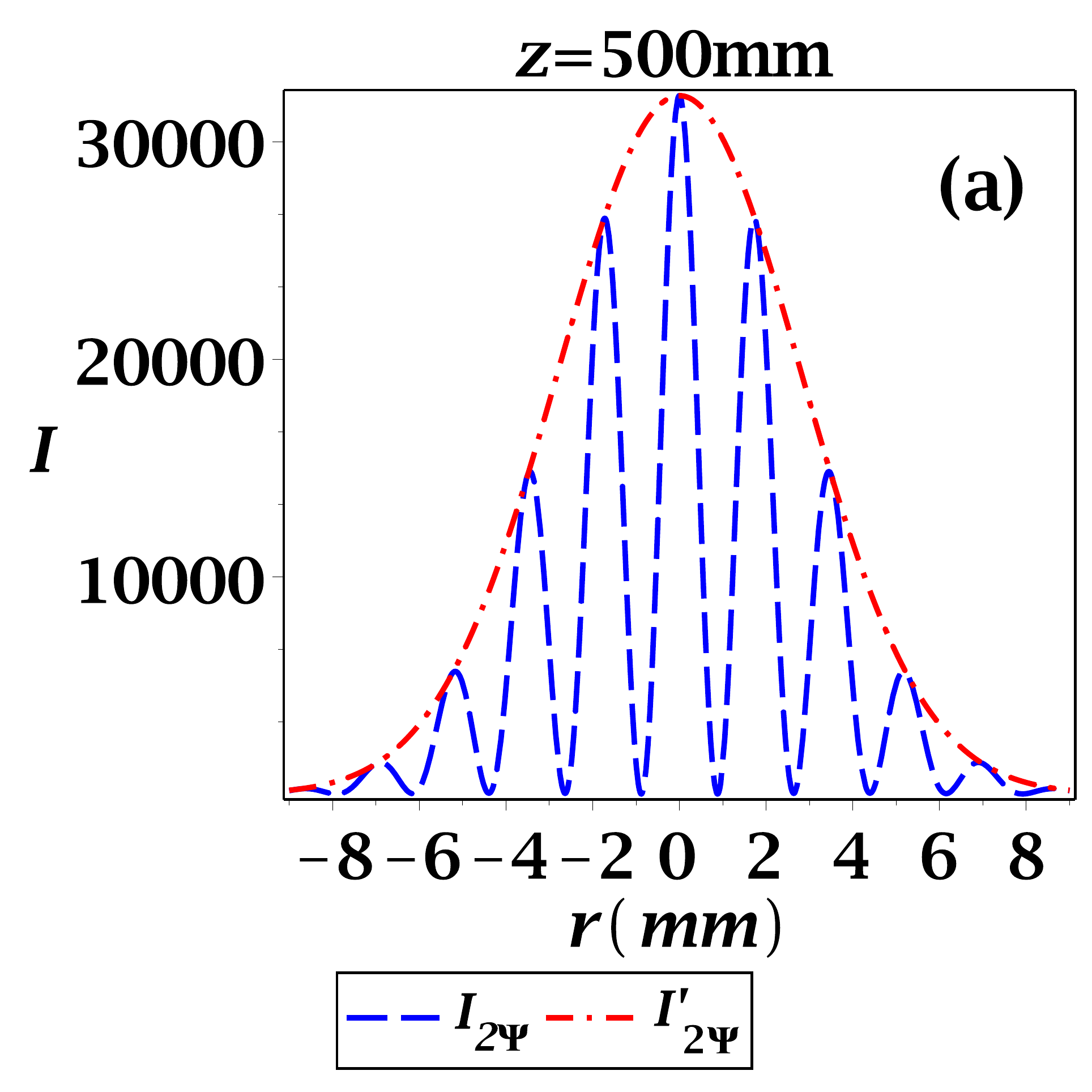}
\includegraphics[width=4.2cm]{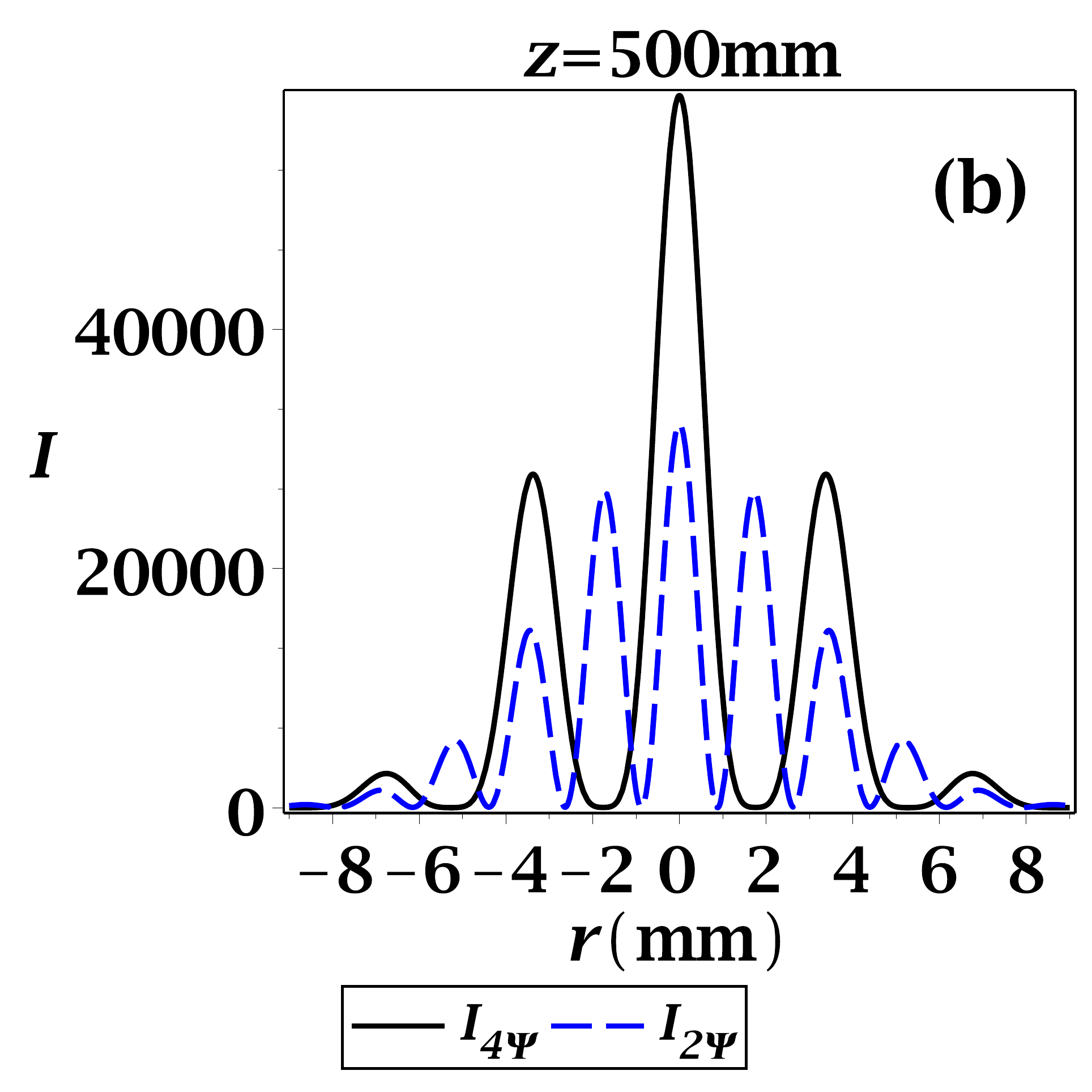}\\
\includegraphics[width=4.2cm]{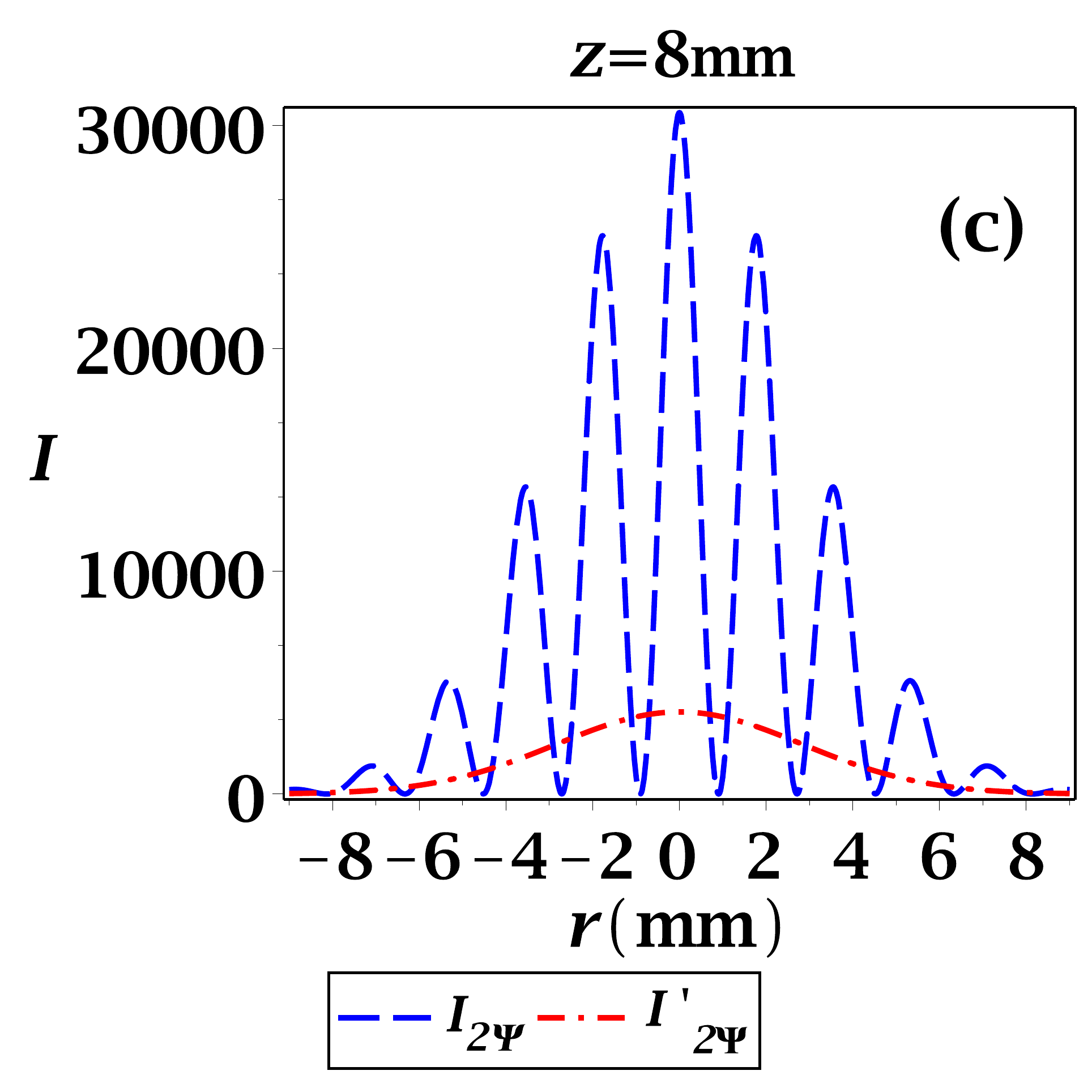}
\includegraphics[width=4.2cm]{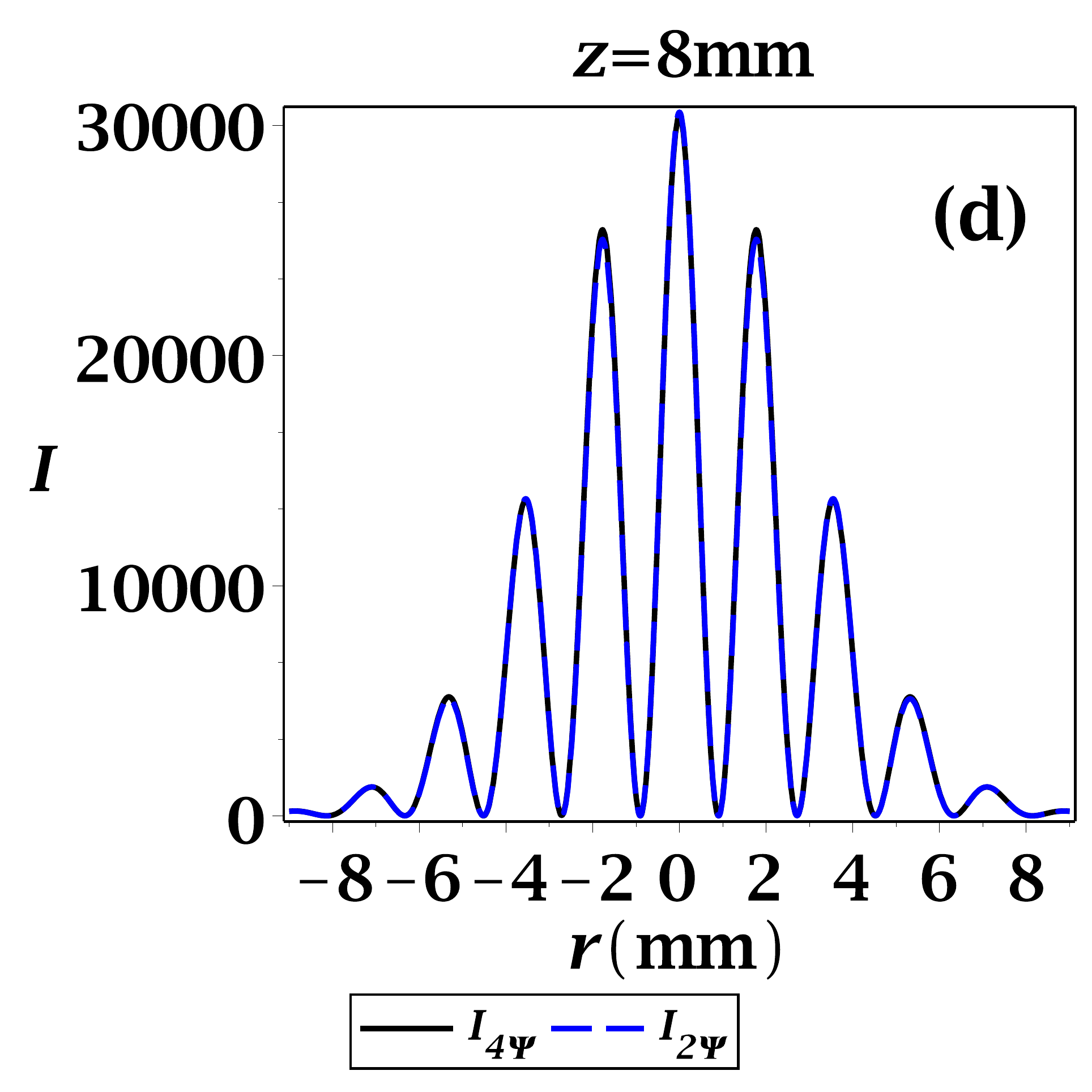}
\caption{
Interference pattern for the biphoton as a function of the transversal position on the detector $x$. 
(a) The dashed line represents the intensity considering only two wavefunctions that describes propagation through the same slit. The dash-dotted line exhibits the intensity considering only photons going through different slits. 
(b) The solid line represents the intensity taking into account the four possibilities of propagation, while the dashed line describes the intensity for the propagation through the same slit.
For the upper plots, the position cross-correlation for the photons just after leaving the slits are $\rho_{uu}=0.98$ and $\rho_{ud}=-0.98$; at the detector placed in $z_{\tau}=500\;\mathrm{mm}$ we have  $\rho_{uu}=0.00035$ and $\rho_{ud}=0.000035$, with $z=500\;\mathrm{mm}$. 
(c) The dashed line represents the intensity for the photons passing through the same slit, and the red dash-dotted line exhibits the intensity considering only two propagations for each photon crossing different slits. 
(d) The solid line represents the intensity taking into account the four possibilities for the propagation, while the dashed line describes the intensity for the photons passing through the same slit. 
For the lower plots, the photons have position cross-correlation $\rho_{uu}=0.98$ and $\rho_{ud}=-0.98$ just after the slits, whereas far from the slits $\rho_{uu}=-0.0099$  and $\rho_{ud}=-0.010$, considering $z=8\;\mathrm{mm}$ and $z_{\tau}=500\;\mathrm{mm}$, respectively.}
\label{intensidades}
\end{figure}
\begin{figure}[htp]
 \centering
\includegraphics[height=4.5cm,width=4.1 cm]{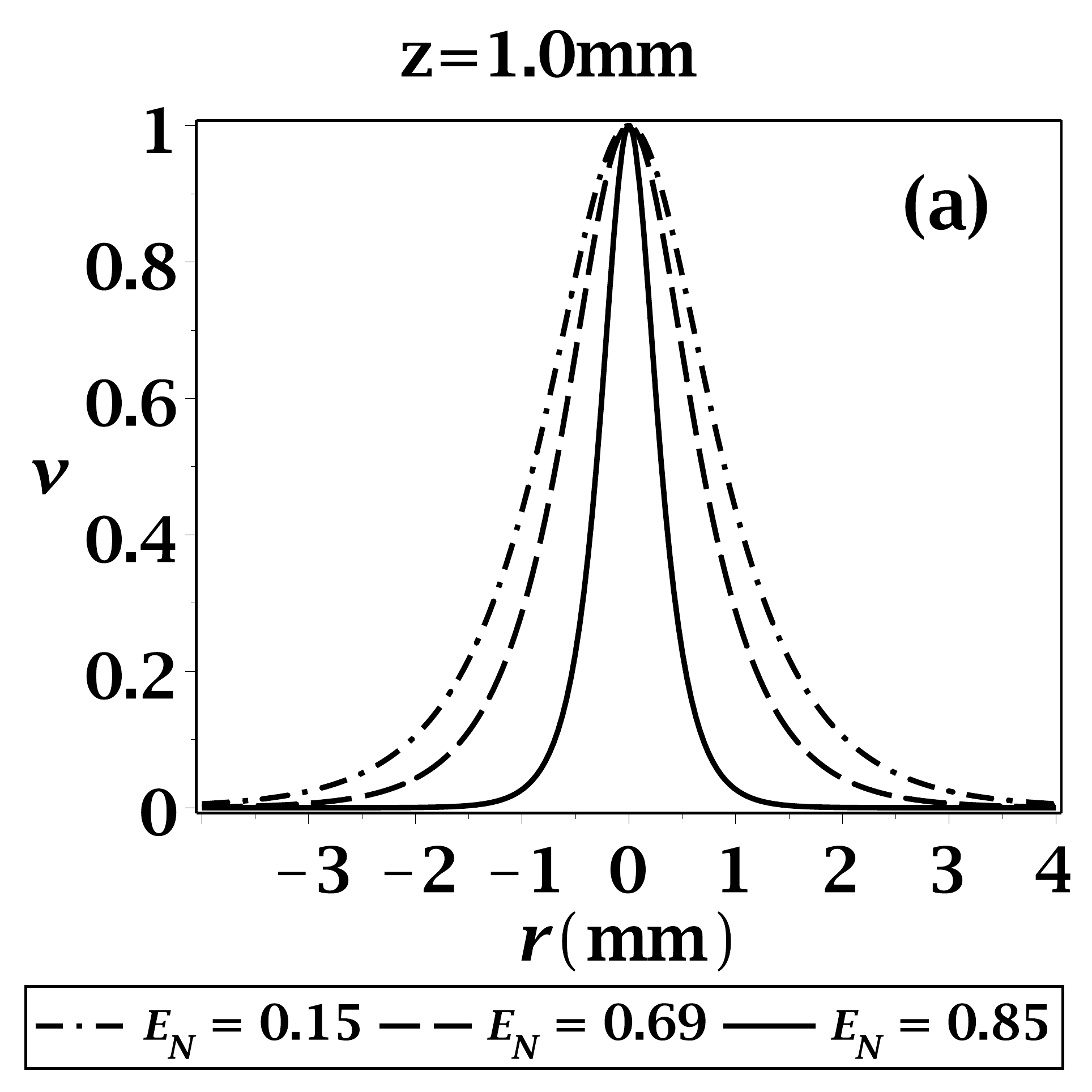}
\includegraphics[height=4.5cm,width=4.1cm]{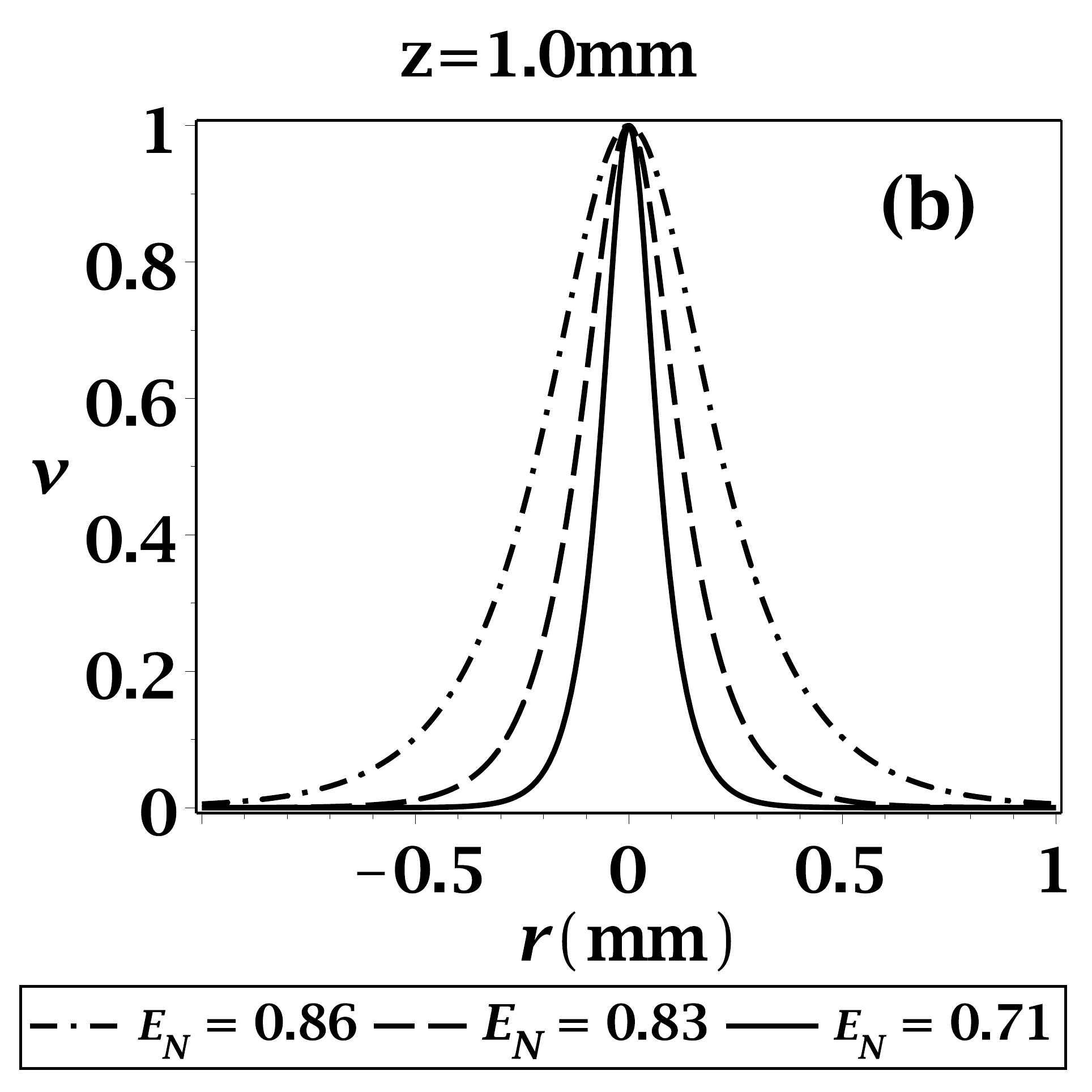}\\
\includegraphics[height=4.1cm,width=5 cm]{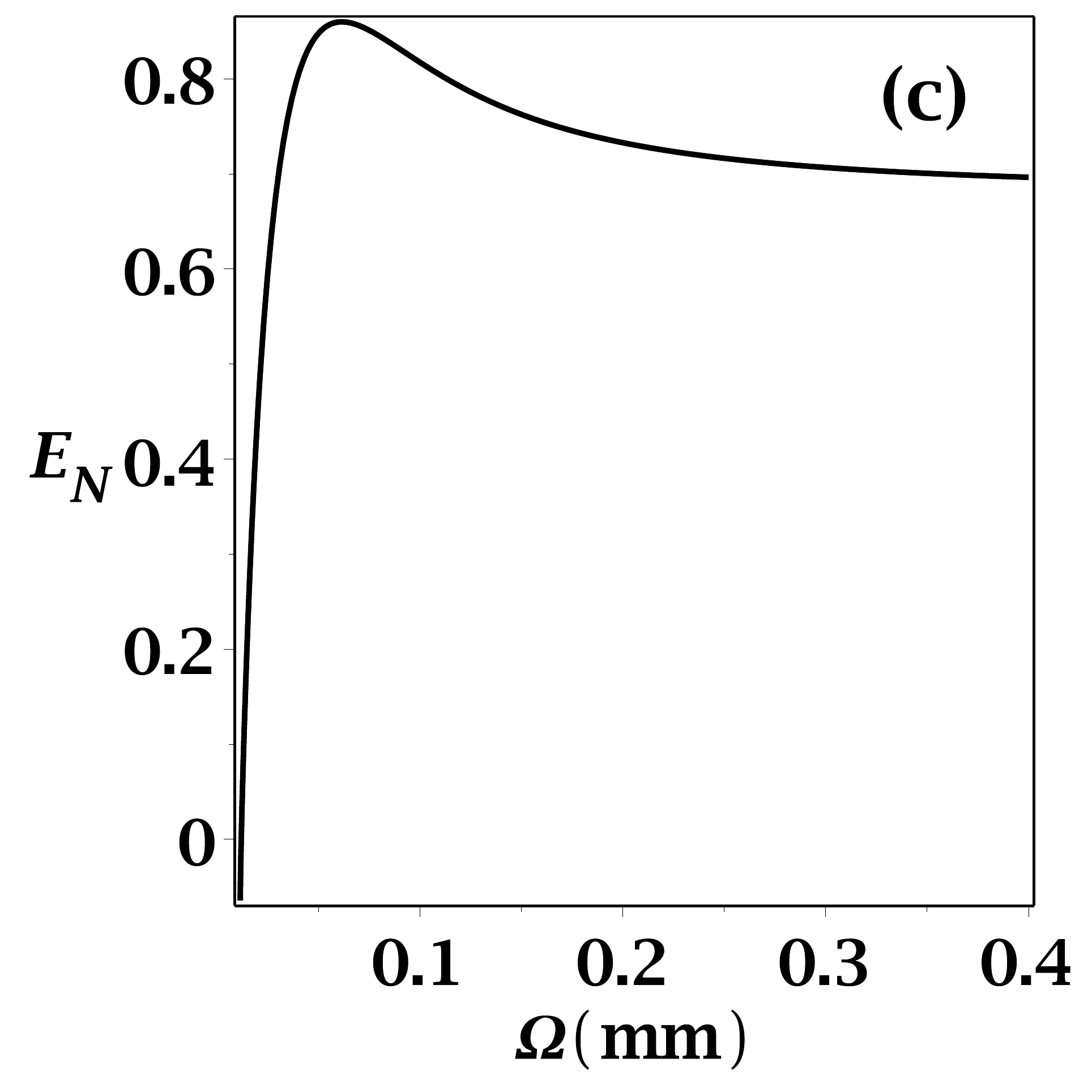}
\caption{The visibility $\nu$ has a maximum value $\nu =1$ at the center $x=0$ independent of the behavior of the logarithmic negativity $E_N$ equation (\ref{negatividade0}) at the detection screen, which is a characteristic of a totally coherent system. On the other hand, outside the center $r\neq0$, the visibility is influenced by the behavior of the entanglement $E_N$. (a)  The visibility $\nu$ becomes less wide as the logarithmic negativity increases, see the dash-dotted, dashed and solid curves, respectively, for $E_N=0.15$, $E_N=0.69$ and $E_N=0.85$.  (c) The visibility $\nu$ is wider for larger values of logarithmic negativity, for instance, $E_N=0.86$,
$E_N=0.83$ and $E_N=0.71$, dash-dotted, dashed and solid curve, respectively. 
 As can be seen in (c), there  are two regions of $\Omega$, where the $E_N$ increases and decreases,  which marks the same behavior for $\nu$ in the increasing and decreasing range of $E_N$ values.}
\label{vis_neg}
\end{figure}

We calculated the intensities $I_{4\psi}$, $I_{2\psi}$ and $I^{\prime}_{2\psi}$ as a function of the position at the screen $x$  as shown in figure \ref{intensidades}. We consider the set of  parameters $\sigma=11.4\;\mathrm{\mu m}$, $\Omega=10\sigma$, $d=100\;\mathrm{\mu m}$, $\beta=5.0\;\mathrm{\mu m}$ and $z_{\tau}=500\;\mathrm{mm}$. The flight distance before the slits is $z=500\;\mathrm{mm}$ for the upper graphs, and $z=8\;\mathrm{mm}$ for the lower graphs. 
The interference fringes produced by $I_{2\Psi}$ (dashed curve in figure \ref{intensidades}(a)) are composed by the  wavefunctions corresponding to the propagation of the photons through the same slit (strongly correlated at the slit). 
On the other hand the wavefunctions corresponding to the propagation of photons through different slits (photons anti-correlated at the slits) produce only the envelope $I'_{2\Psi}$ given by the dash-dotted curve for the interference pattern.  
In figure \ref{intensidades}(b), we compare the intensity $I_{4\Psi}$ (solid curve) that contains the four amplitudes for the biphoton diffraction with $I_{2_\Psi}$ (dashed curve) using the same parameters as those used in figure  \ref{intensidades}(a). 
As we can observe, the interference pattern produced by the four wavefunctions has a few interference fringes as compared to $I_{2\Psi}$, which means that the wavefunctions in $I'_{2\Psi}$ are still giving relevant contribution to $I_{4\Psi}$. In other words, we conclude that a stronger position correlation between the photons  play an important role to unravel the effective contributions to the interference pattern.
Notice that we have the behavior shown in figures \ref{intensidades}(a) and \ref{intensidades}(c), the wavefunctions in $I'_{2\Psi}$ have an even smaller contribution. This means that its wavefunctions contribute less to $I_{4\Psi}$. In fact by comparing the intensities  $I_{4\Psi}$ (solid curve) and  $I_{2\Psi}$  (dashed curve) in figure \ref{intensidades}(d), one notices that its interference patterns match. Thus, the wavefunctions present in $I_{2\Psi}$ govern the interference pattern. Meaning that, under this choice of parameters that leads to  strong position correlation for the biphoton at the slit and anti-correlation at the detection screen, it is reasonable to take into account only the wavefunctions $\Psi_{uu}$ and $\Psi_{dd}$.
In such a regime, the relative intensity $I_r=I/F$ for a biphoton is given by \cite{bramon}
\begin{equation}
I_r(r)=\big[1+\nu (r)\cos\phi (r)\big], \label{intrelclas}
\end{equation}
where, $F(r)=|\Psi_{uu}|^2+|\Psi_{dd}|^2$ and
\begin{equation}
\phi=\phi_{uu}-\phi_{dd}=\left( \frac{ k_0}{cR_{+}}-\frac{
k_0}{cR_{-}} \right)r^2,
\end{equation}
where, $R_{\pm}$ is the radius of curvature of the wave fronts for the propagation through the slit (whose expression is in Appendix \ref{appds}). We have fixed $q=0$, i.e., the two photons strike at the same
position on the screen. From equation (\ref{intrelclas}), the visibility is defined as
\begin{equation}\label{visibilidade1}
\nu(r)=\frac{2|\Psi_{uu}||\Psi_{dd}|}{|\Psi_{uu}|^2+|\Psi_{dd}|^2}=\cosh^{-1}\left(\frac{2
D_{uu} r}{B^2}\right),
\end{equation}
where, $D_{uu}$ and $B$ are the wavepacket separation and wavepacket  spreads, respectively (see Appendix \ref{appds}).
The visibility equation (\ref{visibilidade1}), using the logarithmic negativity at the screen equation (\ref{negatividade0}), as a function of the detector position $r$ and for different values of the logarithmic negativity $E_N$ is exhibited in figures \ref{vis_neg}(a)  \ref{vis_neg}(b) and \ref{vis_neg}(c), where the last represents the logarithmic negativity as a function of an initial entanglement parameter $\Omega$. We use the following set parameters: $\lambda=702\;\mathrm{nm}$,
$\lambda_p=351.1\;\mathrm{nm}$, $L_z=7.0\;\mathrm{mm}$,
$\sigma=\sqrt{\frac{L_p \lambda_p}{6 \pi}}=11.4\;\mathrm{\mu m}$,
$d=200\;\mathrm{\mu m}$, $\beta=60\;\mathrm{\mu m}$,
$z=1\;\mathrm{mm}$ and $z_{\tau}=70\;\mathrm{mm}$.

For a set of three increasing (a) and three decreasing (b) values of logarithmic negativity in figure \ref{vis_neg}, we obtain the same behavior for the visibility. In order to understand this apparent inconsistency, we investigate the logarithmic negativity as a function of the parameter $\Omega$ (c).  Up to $\Omega\approx0.05\;\mathrm{mm}$, the logarithmic negativity $E_N$ increases monotonically, and then decreases down to a constant value. This explains why in (a) the visibility becomes wider both when the logarithmic negativity increases and decreases (b). For $E_N=0.15$, $E_N=0.69$ and $E_N=0.85$, $E_N$ grows in (c) for $\Omega=0.01\;\mathrm{mm}$, $\Omega=0.03\;\mathrm{mm}$ and
$\Omega=0.05\;\mathrm{mm}$, respectively. On the other hand for  $E_N=0.86$,
$E_N=0.83$ and $E_N=0.71$, we have respectively $\Omega=0.06\;\mathrm{mm}$, $\Omega=0.09\;\mathrm{mm}$ and  $\Omega=0.28\;\mathrm{mm}$.

The visibility of the biphoton in a double slit has a maximum value $\nu =1$ at the center $r=0$
regardless of the value of $E_N$ at the detection screen, which is a characteristic of a totally coherent system.
On the other hand, outside the region  $r\neq0$, the visibility is influenced by the behavior of the entanglement. In fact, for the double-slit parameters chosen here,  outside the detector center $r\neq0$, the visibility can cover a larger domain depending on the logarithmic negativity $E_N$.

\section{Gouy phase difference and logarithmic negativity for a type-I SPDC  biphotons}
\label{SectionIII}

For a Type-I SPDC biphoton diffraction through a double-slit, we have shown that one can access information about the logarithmic negativity through values of the Gouy phase (see the lower plot in figure \ref{gouy_N_beta}). 
However, only a Gouy phase difference can be measured in a double-slit experiment. 
The Gouy phases acquired by the biphoton propagating through the upper and lower slits are equal if the slits have the same width.  Thus, in order to measure a Gouy phase difference at the detector, one has to consider an asymmetric, namely different slit widths, double slit setup. 
The Gouy phase difference can be assessed by measuring the relative intensity and the fringe visibility in an asymmetric double-slit. 
In the previous section, we have seen that for a judicious choice of position correlations through a set of parameters, the contribution of two wavefunctions  is suppressed in the intensity which allows us to define the intensity and fringe visibility just as for the case of single particle interference. 

\begin{figure}[ht!]
 \centering
\includegraphics[height=4.5cm,width=5 cm]{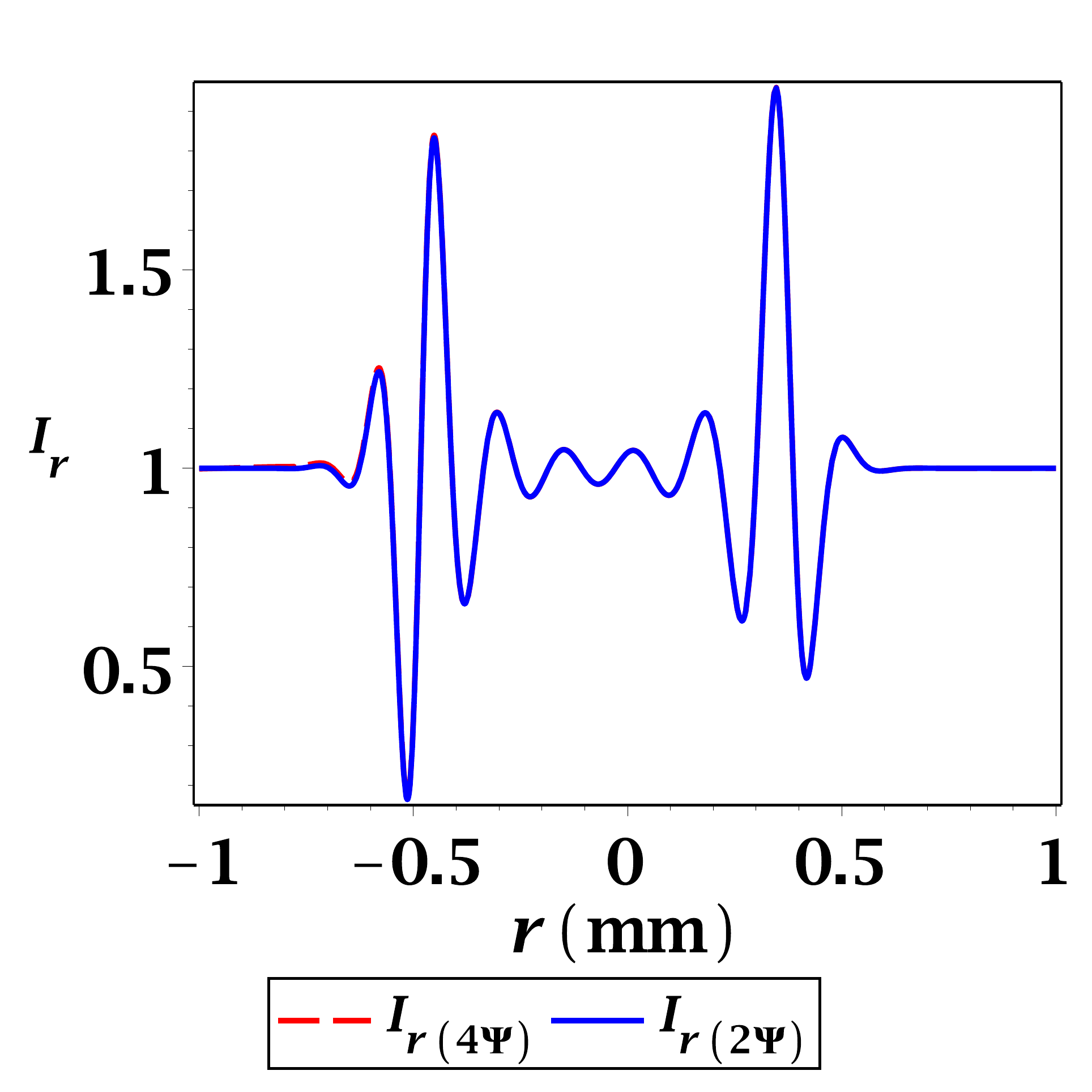}
\includegraphics[height=4.5cm,width=5cm]{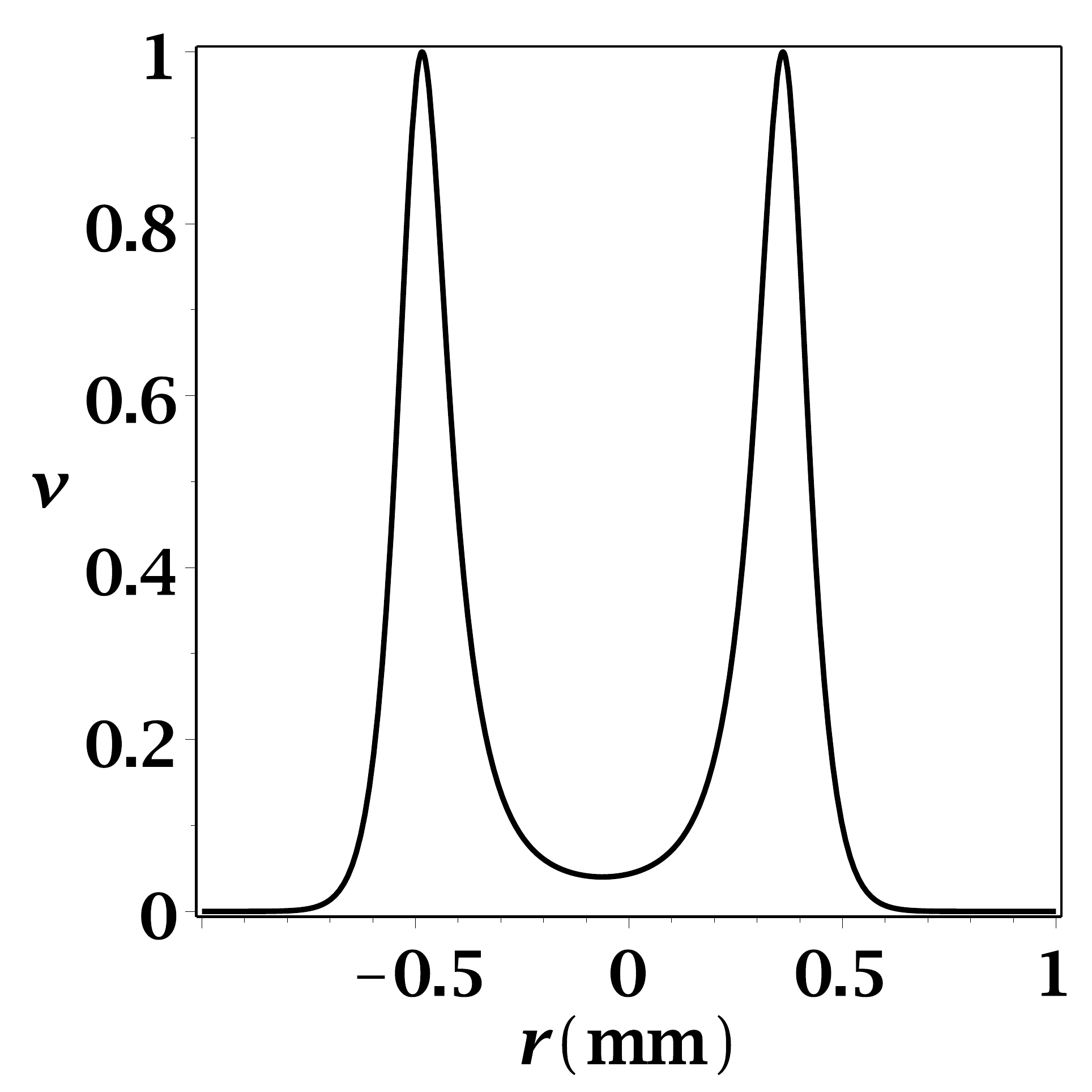}
\caption{ (Top) The relative intensity $I_r$, as a function of the position $r$, considering the four wavefunctions for all  possibilities for the photons to cross the double-slit is represented by the dashed curve $I_{r (4\psi)}$. The relative intensity taking into account only  two wavefunctions for the photons going through the same slit is represented by the solid curve $I_{r (2\psi)}$. We can see that the two curves are approximately equal, so even for different slit widths, we can still find a regime of parameters where two wavefunctions can be safely neglected. (Bottom) The visibility $\nu$ as a function of the transversal position at the screen $r$ for an asymmetric double-slit experiment with biphotons. As a consequence of the different slit width, the visibility $\nu$ presents two maxima which are shifted from the center of the position at the screen $r=0$.}
\label{ir_vis}
\end{figure}

From the equation (\ref{intrelclas}), the visibility is written as
\begin{equation}
\begin{split}
\nu(r)=& \frac{I_r-1}{\cos\big[\phi(r)\big]},
\end{split}\label{visibilidade}
\end{equation}
and since we are considering an asymmetric double-slit, the phase
\begin{equation}
\phi=(\phi_{uu}-\phi_{dd})+(\zeta_{1}-\zeta_{2})
\end{equation}
has a contribution from the Gouy phase difference ($\zeta_{1}-\zeta_{2}$). Notice that the phase difference
\begin{equation}\label{difuudd}
\phi_{uu}-\phi_{dd}=\left( \frac{ k_0}{cR_{+1}}-\frac{ k_0}{cR_{-2}}
\right) {r}^{2}+ \left( \Delta_{1}+\Delta_{2} \right)
r+(\theta_1-\theta_2),
\end{equation}
is given in terms of $\Delta_{1,2}$ and $\theta_{1,2}$. These new contributions stem from different slit widths. The label 1 (2) refers to the upper (lower) slit, and we have fixed $q=0$, i.e. the two photons strike the same position  at the screen. 
 The Gouy phase difference can be obtained from the equation
\eqref{visibilidade}, if the phase difference
$(\phi_{uu}-\phi_{dd})=n\pi$ rad (where $n=2, 4, 6,...$). In this way, the only phase difference that contributes to the relative intensity is the Gouy phase difference. Since, the phase difference $(\phi_{uu}-\phi_{dd})$ depends on the setup parameters, one can easily fulfill the requirement. The Gouy phase difference can be expressed  as 
\begin{equation}
(\zeta_1-\zeta_2)=\arccos\bigg[\frac{I_r(r)-1}{\nu(r)}\bigg].
\label{difz}
\end{equation}
If one has access to the relative intensity $ I_r$ and to the fringe visibility $\nu(r) = (I_{max} -I_{min}) /(I_{max} +I_{min})$ in a double-slit experiment, the Gouy phase difference can be measured. The relative intensity $I_r$ as a function of the position $r$ (exhibited in the upper plot of figure \ref{ir_vis}) for the four biphoton wavefunction contributions (represented by the dashed curve) matches $I_r$ using two wavefunctions only (solid curve), representing the biphotons crossing the upper or lower slits. 
Thus, using an asymmetric double-slit, we can neglect two wavefunctions, without  significant loss of accuracy in the description of our system.  
The behavior of the visibility $\nu$ at the detection screen for  photons passing through slits with different widths, can be seen in the lower plot of figure \ref{ir_vis}. 
As a consequence of taking different slit widths,  the visibility presents two maxima which are shifted from
the center $r=0$. We consider the following set values of
parameters in figure \ref{ir_vis}: $\lambda=702\;\mathrm{nm}$,
$\lambda_p=351.1\;\mathrm{nm}$, $L_z=7.0\;\mathrm{mm}$,
$\sigma=\sqrt{\frac{L_p \lambda_p}{6 \pi}}=11.4\;\mathrm{\mu m}$,
$z_{0-}=k_0\sigma^{2}=1.4\;\mathrm{mm}$, where $k_0=2\pi/ \lambda$,
$\Omega=10\sigma$, $d=200\;\mathrm{\mu m}$, $z=2\;\mathrm{mm}$,
$z_{\tau}=70\;\mathrm{mm}$, $\beta_{1}=60\;\mathrm{\mu m}$ and
$\beta_{2}= 5\;\mathrm{\mu m}$. 

In order to single out the Gouy phase difference in the relative intensity, we have chosen $(\phi_{uu} - \phi_{dd} = n \pi$, where $n= 2, 4, 6, ...$). This automatically selects the set of parameters that fulfill this constraint. Notice that the inter-slit distance $d$ also appears in the definition of the Gouy phase difference written in equation (\ref{difz}) (as it does in the logarithmic negativity) through the parameter $D_{uu}$ contained in the visibility equation (\ref{visibilidade1}). However, the condition  $(\phi_{uu} - \phi_{dd} = n \pi$, where $n= 2, 4, 6, ...$) eliminates automatically the dependence on $d$ in the Gouy phase difference, which comes only from the terms expressed in equation (\ref{difuudd}). 

As a next step, consider the same parameters as used in figure \ref{ir_vis}, except for the  slit widths  $\beta_1$ and $\beta_2$. 
 In order to have $(\phi_{uu}-\phi_{dd})=n\pi$ (where $n=2, 4, 6,...$), we set a value of $\beta_1$, which in turn requires  a specific value of the position $r$ at the detection screen (see equation (\ref{difuudd})).
 For each value of $\beta_1$, there is one correspondent specific value of the Gouy phase difference $|\zeta_1-\zeta_2|$, which is obtained from equation (\ref{difz}) by measuring the relative intensity and the visibility. 
 The corresponding values of $r$, $\beta_1$ and $|\zeta_1-\zeta_2|$ are shown in Table I. 
 Thus, the study of  diffraction of a biphoton through its effective wavefunction in a double-slit  enables us to measure the Gouy phase difference.

\begin{table}[!htb]
\caption{Gouy phase difference as a function of the slit width $\beta_1$.} 
\centering

  \begin{tabular}{c c c c} \hline  \hline \tabularnewline

\hspace*{0.2cm}   $r ~ (\mathrm{mm})$ &\hspace*{0.3cm}  $\beta_1 ~ (\mathrm{\mu m})$ &\hspace*{0.5cm}  $|\zeta_{1}-\zeta_{2}|(rad)$ & \hspace{0.3cm} $E_N =\frac{|\zeta_2-\zeta_1|- 0.48}{0.16}$\vspace{0.2cm}  \tabularnewline \hline 
\addlinespace[3pt]
$-0.123$&  ~~~       $10 $&  $0.105$& $-$ \\
\addlinespace[3pt]
$-0.123$&  ~~~       $15 $&  $0.220$ & $-$ \\
\addlinespace[3pt]
$-0.123$&  ~~~       $20 $&  $0.315$& $-$\\
\addlinespace[3pt]
$-0.124$&  ~~~       $30 $&  $0.436$ & $-$\\
\addlinespace[3pt]
$-0.124$&  ~~~       $36 $&  $0.486$ & $0.0254$  \\
\addlinespace[3pt]
$-0.124$&  ~~~       $40 $&  $0.515$ & $0.206$ \\
\addlinespace[3pt]
$-0.125$&  ~~~       $45 $&  $0.548$&  $0.410$ \\
\addlinespace[3pt]
$-0.125$&  ~~~       $50 $&  $0.578$ &$0.598$\\

\hline\hline
\end{tabular}
\label{tabeladifmu}
\end{table}

Let us summarize our conclusions. The logarithmic negativity $E_N$ is an entanglement quantifier, which in the case of a double-Gaussian describing the biphoton, can be calculated through its covariance matrix. 
It is known that the Gouy phase shift is given in terms of the covariance matrix elements as well \cite{feng2001}. 
In the experiment reported in  \cite{Kawase}, the quantum correlations of a biphoton free evolution are related to their Gouy phase. Likewise, we obtain a relation between these two quantities for the case of a biphoton diffracting through a double-slit.

That being said, two crucial issues appear in this proposal. First, the primarily measurable quantities in a double-slit experiment are the relative intensity and the visibility of the interference fringes. 
Secondly, the logarithmic negativity considered refers to the propagation through a single slit, and thus it is not directly related to the measurable quantities in a double-slit setup. It is exactly in this point that the Gouy phase difference plays a fundamental role. 
Because the Gouy phase difference for slits with the same width is zero, we may employ slits with different apertures $\beta_1$ and $\beta_2$. 
Therefore, we may fix $\beta_2$ and vary $\beta_1$ in order to relate the Gouy phase difference (encoded in $\beta_1$) and the logarithmic negativity (supposing that it was calculated for the wavefunction corresponding to slit $1$, that is to say  through $\psi_{uu}$).

\begin{figure}[ht!]
 \centering
 \includegraphics[height=5cm,width=4.5 cm]{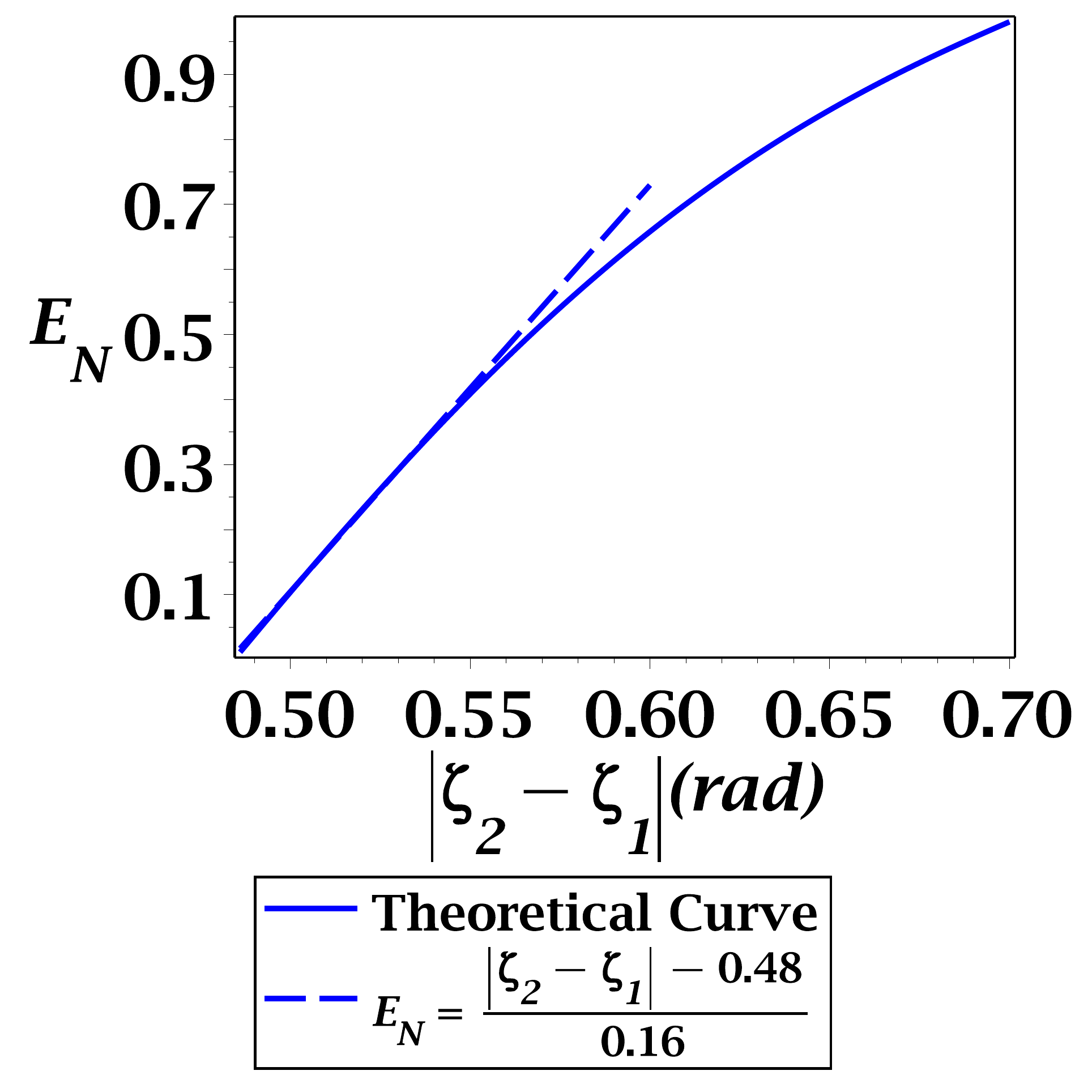}
 \includegraphics[height=4.5cm,width=4.3 cm]{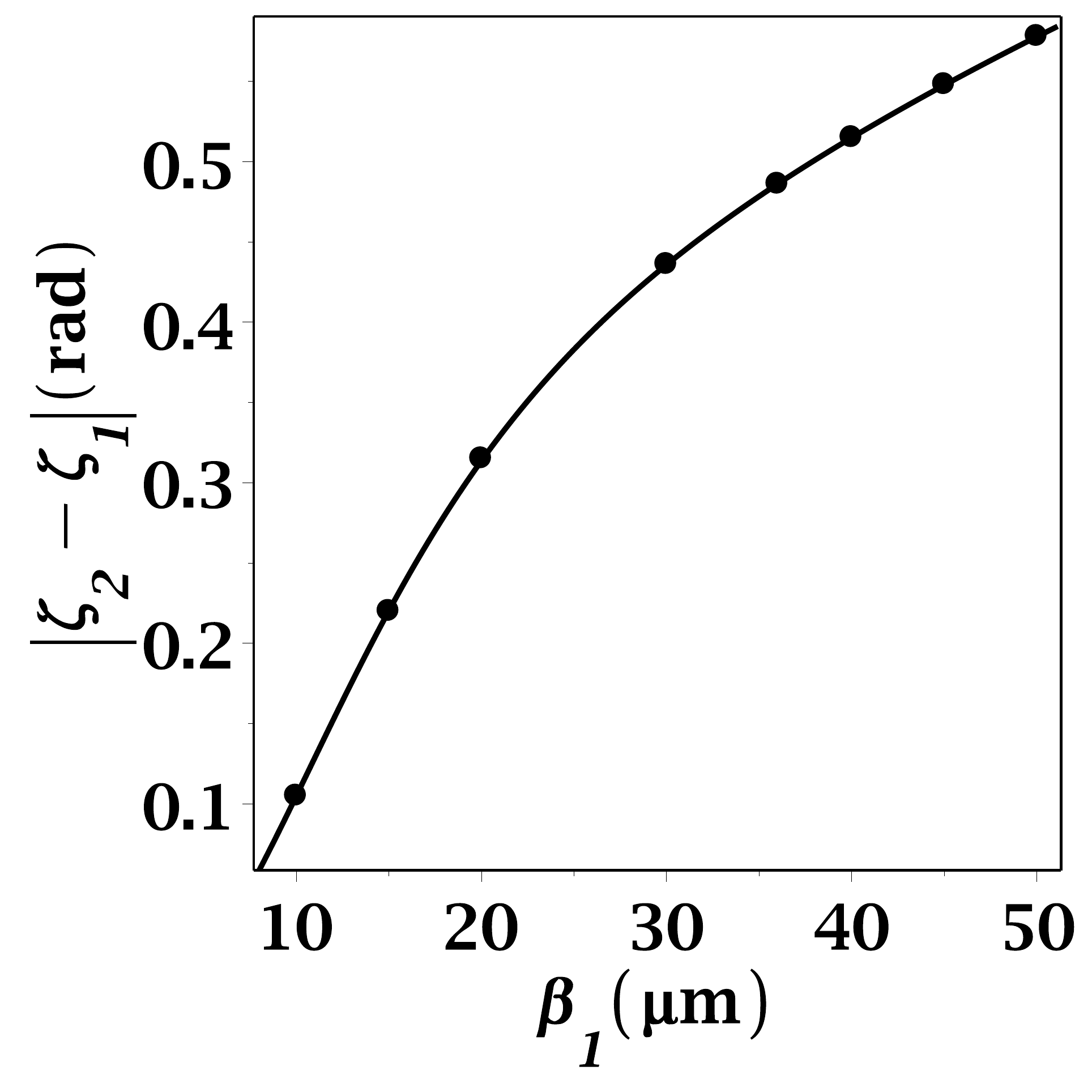}
\includegraphics[height=4.5cm,width=4.2 cm]{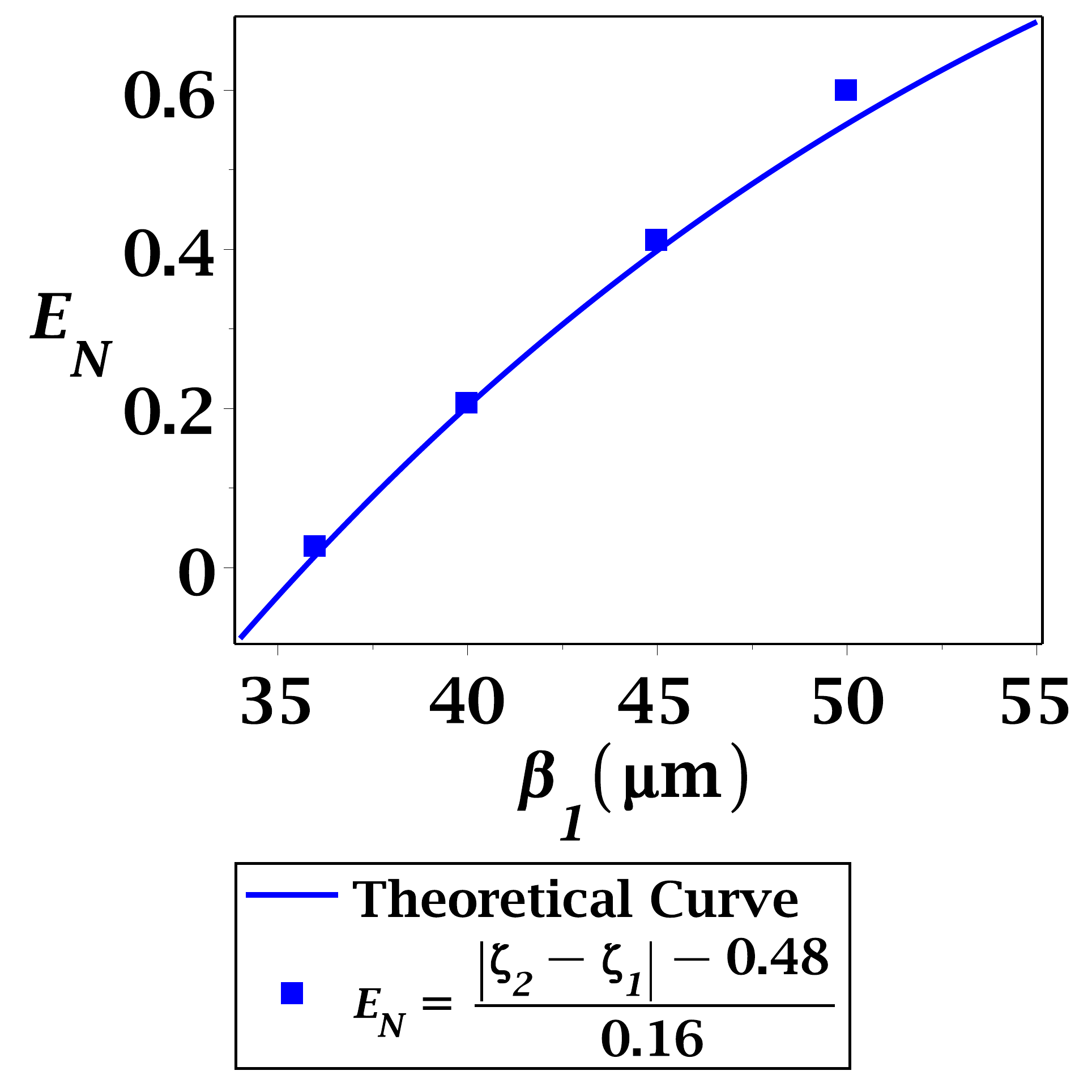}
\caption{(Upper) We can observe how the logarithmic negativity $E_N$ behaves in terms of the  Gouy phase difference  substituting $|\zeta_2-\zeta_1|$ in terms of $\beta_1$ into the logarithmic negativity expression equation (\ref{negatividade0}), which gives us the solid curve. We have observed that the behavior of $E_N$ for a small variation of Gouy phase difference can be reproduced by a linear relation, for instance, $E_N =\frac{|\zeta_2-\zeta_1|- 0.48}{0.16}$ (represented by the dashed curve) describes the behavior in the range of $36\;\mathrm{\mu m}<\beta_{1}<50\;\mathrm{\mu m}$.
(Lower-left) Gouy phase difference $|\zeta_2-\zeta_1|$ as a function of the slit width $\beta_1$: the solid curve is the theoretical curve, and the dotted curve was obtained numerically calculating the relative intensity and the fringe visibility for the parameters that satisfy $(\phi_{uu}-\phi_{dd})\approx n\pi$ (where $n=2, 4, 6,...$) (see Table I). (Lower-right) Logarithmic negativity $E_N$ as a function of the slit width $\beta_1$: the solid curve is the theoretical curve, and the curve with boxes is the negativity values obtained from $E_N =\frac{|\zeta_2-\zeta_1|- 0.48}{0.16}$,  for each value of $|\zeta_2-\zeta_1|$ as a function of $\beta_1$ in Table \ref{tabeladifmu}.
The slit width is what essentially determines the variation of the logarithmic negativity, just as for the Gouy phase difference. 
Thus, we propose to access values of $E_N$ through the slit width from $|\zeta_2-\zeta_1|$ measurements in terms of $\beta_1$.
}\label{pontosgouy}
\end{figure}

We set all the slit parameters, except one of the slit widths, as mentioned before, so to modify the slit width $\beta_1$ implies to vary the Gouy phase difference, which can be obtained from equation (\ref{gouy_slit}).  
Thus, we can write the slit width $\beta_1$ as a function of $|\zeta_2-\zeta_1|$ and substitute it into $E_N$ equation (\ref{negatividade0}). Therefore, we obtain the relation between the logarithmic negativity and the Gouy phase difference of a biphoton diffracting through a double-slit setup,  which is represented by the solid curve in figure \ref{pontosgouy} (upper plot). We emphasize that such a connection between the logarithmic negativity and the Gouy phase difference is possible for any set of parameters.


 The Gouy phase difference (data in Table \ref{tabeladifmu}), numerically calculated using the relative intensity and the visibility, is illustrated in figure \ref{pontosgouy} (dotted curve in the lower-left plot) as a function of $\beta_1$.
The theoretical curve for the Gouy phase difference, represented by the solid curve, is obtained from equation (\ref{gouy_slit}). For the plots in figure \ref{pontosgouy}, we have used the parameters: $\lambda=702\;\mathrm{nm}$,
$\lambda_p=351.1\;\mathrm{nm}$, $L_z=7.0\;\mathrm{mm}$,
$\sigma=\sqrt{\frac{L_p \lambda_p}{6 \pi}}=11.4\;\mathrm{\mu m}$,
$z_{0-}=k_0\sigma^{2}=1.4\;\mathrm{mm}$, where $k_0=2\pi/ \lambda$,
$\Omega=10\sigma$, $d=200\;\mathrm{\mu m}$, $z=2\;\mathrm{mm}$,
$z_{\tau}=70\;\mathrm{mm}$ and $\beta_{2}=5\;\mathrm{\mu m}$.

A suitable way to relate Gouy phase difference measurements and logarithmic negativity values in terms of the $\beta_1$ (both depicted in figure \ref{pontosgouy} lower-left and lower-right plots), for our experimental setup, is if a linear relation exist between these two quantities at a certain interval of $\beta_1$. 
In fact, there are ranges of $\beta_1$ where the variation of the Gouy phase difference $|\zeta_2-\zeta_1|$ is small, and a linear relation  reproduces the theoretical behavior in the region considered. 
We considered the region where the slit width is $36\;\mathrm{\mu m}<\beta_{1}<50\;\mathrm{\mu m}$, which corresponds to $0.486<|\zeta_2-\zeta_1|<0.578$, the theoretical curve can be approximated by $|\zeta_2-\zeta_1|\approx0.16 E_N + 0.48$.
Thus, it allows us to attain an expression to extract the logarithmic negativity in terms of the slit width, valid in the neighborhood of $36\;\mathrm{\mu m}<\beta_{1}<50\;\mathrm{\mu m}$, from experimental indirect measurements of the Gouy phase difference.

Thus, straightforwardly, each measurement  of $|\zeta_2-\zeta_1|$ obeying the constraint $(\phi_{uu}-\phi_{dd})\approx n\pi$ (where $n=2, 4, 6,...$) as a function of $\beta_1$  in Table \ref{tabeladifmu} gives an associated $E_N$ value, through $E_N =\frac{|\zeta_2-\zeta_1|- 0.48}{0.16}$, related to an equivalent slit width $\beta_1$.
We notice in figure \ref{pontosgouy} (lower-right plot) that the behavior obtained for the logarithmic negativity from these data  (curve with square shaped points) is in good agreement with the theoretical  curve (solid curve), calculated through the symplectic eigenvalue equation (\ref{negatividade0}). 
We have noticed that values of $E_N$ for slit widths $\beta_1<36\;\mathrm{\mu m}$ provide negative values of logarithmic negativity, so although we have Gouy phase difference for other $\beta_1$ values, they are not convenient for this analysis. Thus, we do not present them in the Table, and the plot starts from $\beta_1=36\;\mathrm{\mu m}$. 

We show only some points representing possible indirect  measurements of $|\zeta_2-\zeta_1|$, although one may obtain many other points in this interval of $\beta_1$ (satisfying to $(\phi_{uu}-\phi_{dd})=n\pi$ rad, where $n=2, 4, 6,...$) , which yield many other convenient logarithmic negativity values through $\beta_1$ in terms of $|\zeta_2-\zeta_1|$. Notice that in figure \ref{gouy_N_beta}, the logarithmic negativity decreases as the Gouy phase $\zeta$ increases, in the present case, we have the opposite behavior, one increases as the other one increases as well. However, here we are considering the behavior of the logarithmic negativity \textit{versus} the Gouy phase difference $|\zeta_2-\zeta_1|$, which increases monotonically. 

Therefore, we have achieved our goal of obtaining the quantum correlations behavior, encoded in the logarithmic negativity, in terms of the Gouy phase difference of a pair of entangled photons diffracting through a double-slit. 
 In addition, we considered a region of small variation of this Gouy phase difference as a function of the slit width, where a linear relation between $E_N$ and $|\zeta_2-\zeta_1|$ is capable of reproducing the theoretical curve, allowing one to obtain $E_N$ as a function of $\beta_1$ through indirect measurements of Gouy phase difference in terms of $\beta_1$ as well. Although, the Gouy phase difference variation is small, a large range of logarithmic negativity values is obtained in that interval, as we have a wide range of slit width values.

\section{Conclusions and final remarks}
\label{SectionIV}
We proposed a scheme to measure biphoton spatial correlations 
as it propagates through a double-slit.
Our proposal is based on measuring the biphoton Gouy phase
difference and the logarithmic negativity in the double-slit
experiment. We considered an asymmetric double-slit experiment with
biphotons and calculated the wavefunctions at the detection
screen corresponding to the four possibilities of propagation
through the slits. Next, we calculated the logarithmic negativity at
the detection screen and showed that it is dependent on the initial
entanglement and the geometrical parameters of the double-slit. We found
conditions for which the interference pattern corresponds to the one produced by the two
wavefunctions corresponding to the propagation of photons through
the same slit. These conditions are related with the behavior of the
cross correlations at the double-slit as well as at the detection
screen. Under such conditions we calculated the fringe visibility.
We observed that the behavior of the visibility is influenced by the
behavior of the initial entanglement created by the nonlinear crystal and the entanglement modification imposed by the diffraction of the twin photons through the double-slit, encoded in the logarithmic negativity at the detection screen.
Then we considered slits with different widths in order to
produce a Gouy phase difference. We expressed that Gouy
phase difference in terms of the relative intensity and the
visibility, showing that it can be experimentally assessed.
Finally, we obtained the quantum correlations behavior, encoded in the logarithmic negativity, in terms of the Gouy phase difference (as the logarithmic negativity has dependence on the slit width, which in turn can be written in terms of the Gouy phase difference) of a pair of entangled photons diffracting through a double-slit. Moreover, we found conditions in which a linear relation between the logarithmic negativity and the Gouy phase difference is capable of reproducing the theoretical behavior, allowing one to obtain the logarithmic negativity, as a function of the slit width, from indirect measurements of the Gouy phase difference, in terms of slit width as well. Therefore, measuring the Gouy phase difference allow us for accessing information about quantum correlations at the detection screen as a function of the slit aperture.

Spatially correlated photonic qutrit pairs were proposed theoretically and experimentally tested in \citep{SINHAQT}. Those qutrits were produced by parametric down-converted biphotons passing through a three-slit apparatus and displayed a high spatial correlation, with a Pearson coefficient of about 0.9. Such correlations are governed by geometrical parameters such as  the slit width, the inter-slit distance and the nonlinear crystal longitudinal length, which also characterize the Gouy phases in our model. An important point raised in \cite{SINHAQT} regards the truly quantum nature of qutrit correlations. Our approach using the  relation between the Gouy phase differnce and the logarithmic negativity can be an important tool to evaluate the nature of the continuous variable correlations in this case. Moreover, in \citep{Kawase}, it was proposed theoretically and experimentally that the relative phase of two different Laguerre-Gauss modes of biphotons can be manipulated via the Gouy phase. Their result suggests the Gouy phase as a new tool to manipulate multidimensional photonic quantum states. In our model, we explicitly demonstrate how to measure the Gouy phase difference for a similar system and establish its connection to the quantum correlations as encoded in the logarithmic negativity. In addition, just as in \citep{Kawase}, our scheme is inherently nondestructive as the Gouy phase difference is established by a two-slit interference and assessed via the visibility and the relative intensity as seen in the interference pattern.

\section*{Acknowledgements}
This study was financed in part by the Coordenação de Aperfeiçoamento de Pessoal de N{\'i}vel Superior - Brasil (CAPES) - Finance Code 001. FCVB thank CAPES-PRINT (PRINT - Programa Institucional de Internacionalização) for the grant 88887.574672/2020-00. FCVB is also grateful for the fruitful discussions with Elizabeth Agudelo, and for the provided space by the Institut für Quantenoptik und Quanteninformation Wien, and by the Atominstitut, TU Wien. IGP thanks Conselho Nacional de Desenvolvimento Cient{\'i}fico e Tecnol{\'o}gico (CNPq), Grant No. 307942/2019-8. JBA thanks CNPq for the Grant No. 150190/2019-0, and MS thanks CNPq for the Grant No. 303482/2017.

\appendix
\section{Biphoton free propagation wavefunction Constants}
\label{freeprop}


The photon pair state for a free propagation after a general distance $z$, in terms of the relative coordinates $r=(x_1+x_2)/2$ and $q=(x_1-x_2)/2$, is given by \cite{crisfree}
\begin{equation}
 \begin{split}
\psi(r,q,z)=& \frac{1}{\sqrt{4\pi w(z)\tilde{w}(z)}}\exp \bigg
\lbrace -\bigg[\frac{ r^2}{w^2(z)} +\frac{q^2}{\tilde{w}^2(z)}
\bigg] \bigg \rbrace \\ & \times \exp \bigg \lbrace -i
\bigg[-\frac{k_0}{r_{+}}r^2- \frac{k_0}{r_{-}}q^2+\zeta(z)\bigg]
\bigg \rbrace,
\end{split}
\end{equation}
where, the wavepacket spread for the signal and idler
beams is written as
\begin{equation}
w^2(z)= \Omega^2 \bigg[1+\bigg(\frac{z}{z_{0+}}\bigg)^2\bigg],
\hspace{0.3cm} \tilde{w}^2(z) =\sigma^2
\bigg[1+\bigg(\frac{z}{z_{0-}}\bigg)^2\bigg].
\end{equation}

Respectively, the radius of curvature of the wave fronts and the longitudinal distance are given by
\begin{equation}
r_{\pm}(z)=z \bigg[1+\bigg(\frac{z_{0\pm}}{z}\bigg)^2\bigg],
\hspace{0.5cm}  z=c t.
\end{equation}

Lastly, we can identify the Gouy phase, which is propagation distance dependent, and it carries the parameters of the initial wavepacket, as
\begin{eqnarray}\label{gouy_free}
\zeta(z)&=&-\frac{1}{2} \bigg \lbrace \arctan \bigg[z \bigg(
\frac{z_{0+}+z_{0-}}{z_{0+}z_{0-}-z^2} \bigg)\bigg]\bigg \rbrace,
\end{eqnarray}
where,  the corresponding Rayleigh lengths is

\begin{equation}
z_{0+}=k_0\Omega^2 \hspace{0.2cm}, \hspace{0.2cm} z_{0-}=k_0\sigma^2
\hspace{0.2cm}, \hspace{0.2cm} \text{and} \hspace{0.2cm}
k_0=2\pi/\lambda.
\end{equation}

\section{Biphoton diffracting through a double-slit wavefunction constants}
\label{appds}
The state that describe the twin photons after diffracting a double-slit  is given by 
\begin{equation}
\begin{split}
\Psi_{uu} (r, q)&=\frac{1}{\sqrt{\pi B
\tilde{B}}}\exp\left[-\frac{(r-D_{uu}/2)^2}{B^2}\right]\exp\left[\frac{q^2}{\tilde{B}^2}\right]\\
& \times \exp\left(\frac{ik_0 }{ R_{+}}r^2+\frac{ik_0 }{
R_{-}}q^2+i \Delta_{uu} r + i\theta_{uu}+i\zeta\right),
\end{split}
\end{equation}
where, using the same interpretation as the biphoton free propagation, the wavepacket spreads for the
propagation through the slit are written as

\begin{equation}\label{B}
B^2(z,z_{\tau})=\frac{\left(\frac{1}{\beta^2} +
\frac{1}{w^2}\right)^2+k_0^2\left(\frac{1}{z_{\tau}}+\frac{1}{r_{+}}\right)^2}{\left(\frac{k_0}{z_\tau}\right)^2\left(\frac{1}{\beta^2}
+ \frac{1}{w^2}\right)}
\end{equation}
and
\begin{equation}\label{Btil}
\tilde{B}^2(z,z_{\tau})=\frac{\left(\frac{1}{\beta^2} +
\frac{1}{\tilde{w}^2}\right)^2+k_0^2\left(\frac{1}{z_{\tau}}+\frac{1}{r_{-}}\right)^2}{\left(\frac{k_0}{z_\tau}\right)^2\left(\frac{1}{\beta^2}
+ \frac{1}{\tilde{w}^2}\right)}.
\end{equation}

The radius of curvature of the wave fronts for the propagation through the slit are

\begin{equation}\label{R+}
R_{+}(z,z_{\tau})=z_\tau \frac{\left(\frac{1}{\beta^2} +
\frac{1}{w^2}\right)^2+k_0^2\left(\frac{1}{z_{\tau}}+\frac{1}{r_+}\right)^2}{\left(\frac{1}{\beta^2}+\frac{1}{w^2}\right)^2+\left(\frac{z}{\Omega^2
w^2}\right)\left(\frac{1}{z_{\tau}} + \frac{1}{r_+}\right)}
\end{equation}
and
\begin{equation}\label{R-}
R_{-}(z,z_{\tau})=z_\tau \frac{\left(\frac{1}{\beta^2} +
\frac{1}{\tilde{w}^2}\right)^2+k_0^2\left(\frac{1}{z_{\tau}}+\frac{1}{r_{-}}\right)^2}{\left(\frac{1}{\beta^2}+\frac{1}{\tilde{w}^2}\right)^2+\left(\frac{z}{\sigma^2
\tilde{w}^2}\right)\left(\frac{1}{z_{\tau}} +
\frac{1}{r_{-}}\right)}.
\end{equation}

The wavepacket separation due to the slits is
\begin{equation}\label{Duu}
D_{uu}(z,z_{\tau})=\frac{\left(1+\frac{z_\tau}{r_+}\right)}{\left(1+\frac{\beta^2}{w^2}\right)}d,
\end{equation}
and the phase that plays a role of wavenumber is given by
\begin{equation}\label{deltauu}
\Delta_{uu}(z,z_{\tau})= \frac{z_{\tau} \Omega^2 }{z_{0+}\beta^2
B^2} d.
\end{equation}
where $z_{0\pm}$ is the corresponding Rayleigh lengths. 

The phases dependents of the propagation
distance are
\begin{equation}
\theta_{uu}(z,z_{\tau})=
\frac{d^2}{4\beta^4}\frac{k_0\left(\frac{1}{z_{\tau}}+\frac{1}{r_{+}}\right)}{\left[\left(\frac{1}{\beta^2}+\frac{1}{w^2}\right)^2+k_0^2\left(\frac{1}{z_{\tau}}+\frac{1}{r_{+}}\right)^2\right]}
\end{equation}
and the Gouy phase, identified as

\begin{equation}
\begin{split}
\zeta=& -\frac{1}{2}\arctan\bigg[\frac{f(z,z_\tau, \beta) +g(z,z_\tau, \beta)}{1-f(z,z_\tau, \beta)g(z,z_\tau, \beta)}\bigg],
\end{split}
\end{equation}
where,
\begin{equation}
    f(z,z_\tau, \beta)=\frac{ z+ z_\tau \big( 1+
\frac{\sigma^2}{\beta^2}\big)}{z_{0-} \big(1-\frac{z z_{\tau}
\sigma^2}{z_{0-}^2 \beta^2}\big)}
\end{equation}
and
\begin{equation}
    g(z,z_\tau, \beta)=\frac{ z+ z_\tau \big( 1+
\frac{\Omega^2}{\beta^2}\big)}{z_{0+}\big(1-\frac{z z_{\tau}
\Omega^2}{z_{0+}^2 \beta^2}\big)}.
\end{equation}

Aiming to obtain the expressions for the wavefunction that describes
the two photons propagating through the lower slit,
$\Psi_{dd}(r,q,z,z_{\tau})$, we just have to substitute the
parameter $d$ with $-d$ in the expressions corresponding to the two
photons propagating through the upper slit, i.e., in
$\Psi_{uu}(r,q,z,z_{\tau})$.

Whereas, the wavefunction describing one photon
propagating through the upper slit while the another propagates through the lower slit is given by
\begin{equation}
\begin{split}
\Psi_{ud} (r, q)&=\frac{1}{\sqrt{\pi B \tilde{B}}}\exp\left[-\frac{r^2}{B^2}\right]\exp\left[-\frac{(q-D_{ud}/2)^2}{\tilde{B}^2}\right]\\
& \times \exp\left(\frac{ik_0 }{ R_{+}}r^2+\frac{ik_0 }{
R_{-}}q^2+i \Delta_{ud}q +i\theta_{ud}+i\zeta\right),
\end{split}
\end{equation}
where,
\begin{equation}
D_{ud}(t,\tau)=\frac{\left(1+\frac{z_\tau}{r_-}\right)}{\left(1+\frac{\beta^2}{\tilde{w}^2}\right)}d,\label{Dud}
\end{equation}

\begin{equation}
\Delta_{ud}(t,\tau)= \frac{z_{\tau} \Omega^2 }{z_{0-}\beta^2 B^2} d,
\end{equation}
and
\begin{equation}
\theta_{ud}(t,\tau)=\frac{d^2}{4\beta^4}\frac{k_0\left(\frac{1}{z_{\tau}}+\frac{1}{r_{-}}\right)}{\left[\left(\frac{1}{\beta^2}+\frac{1}{\tilde{w}^2}\right)^2+k_0^2\left(\frac{1}{z_{\tau}}+\frac{1}{r_{-}}\right)^2\right]}.
\end{equation}

In order to obtain $\Psi_{du}(r,q,z,z_{\tau})$, we need to replace $d$ with $-d$ in $\Psi_{ud}(r,q,z,z_{\tau})$. 

\section{Logarithmic negativity and Covariance Matrix terms for a Type-I SPDC pair of photons diffracting in a double-slit}
\label{covmat}
The logarithmic negativity in equation (\ref{negatividade0}) is given by
\footnotesize
\begin{equation}
\begin{split}
E_N=\ln \left[ \frac{\sqrt{2}icB\tilde{B}R_+R_-}{ \sqrt{\sqrt{-A_1A_2}+A_3}}\right], 
\end{split}
\end{equation}
where $i$ is the imaginary unit and $c$ is the speed of light constant. $B$,  $\tilde{B}$, $R_+$, $R_-$ are shown in equation (\ref{B}), (\ref{Btil}), (\ref{R+}) and (\ref{R-}), respectively. In the following,
\begin{equation}
    A_1=\bigg(B^2+\frac{\alpha_1+\alpha_3}{2(\alpha_2+\alpha_4)}\bigg)^2(\alpha_2+\alpha_4)-\frac{(\alpha_1+\alpha_3)^2}{4(\alpha_2+\alpha_4)} + \frac{\alpha_4}{\Delta_{uu}^{2}}(\tilde{B}+D_{uu}^2),\nonumber
\end{equation}
in order to obtain $A_2$, replace $\alpha_1$ and $\alpha_2$ with $-\alpha_1$ and $-\alpha_2$, respectively. Next, we have
\begin{equation}
    A_3=4\alpha_5\bigg(B^2+\frac{\alpha_6}{2\alpha_5}\bigg)^2-\bigg(\frac{{\alpha_6}^2}{\alpha_5}+\frac{8\alpha_4}{\Delta^2_{uu}}\bigg),\nonumber
\end{equation}
where
\begin{equation}
    \alpha_1=\left(  \left( {\it R_-}\,c\Delta_{uu}-{\it D_{uu}}\,{\it k_0} \right) {\it R_+
}+{\it D_{uu}}\,{\it R_-}\,{\it k_0} \right)^{2}{{\it \tilde{B}}}^{4}+{c}^{2}{{
\it D_{uu}}}^{2} R^2_+ R^2_-,\nonumber
\end{equation}
\begin{equation}
    \alpha_2={c}^{2}R^2_+R^2_-+ k_0^2 \left( {\it R_+}-{\it R_-} \right) ^{2}{{\it \tilde{B}}}^{4}, \nonumber
\end{equation}
\begin{equation}
    \alpha_3=2c^2 \tilde{B}^2 R^2_+ R^2_-,
 \hspace{0.7cm}
    \alpha_4=\frac{\alpha_3\Delta^2_{uu}}{2}, \nonumber
\end{equation}
\begin{equation}
   \alpha_5= -2 k_0^2\left( R_+ -R_-\right)^2 {{\it \tilde{B}}}^{4}-2\,{c}^{2}R^2_+R^2_-\nonumber
\end{equation}
and
\begin{equation}
\begin{split}
    \alpha_6=&\left( -2\,{c}^{2}{\Delta_{uu}}^{2}{{\it R_-}}^{2}-4\,c{\it k_0}\,{\it D_{uu}}\,{\it R_+}
\,\Delta_{uu}-2\,{{\it D_{uu}}}^{2}{{\it k_0}}^{2} \right)  \\ & \times\left( 
{\it R_-}-{\frac {\,{\it k_0}\,{\it R_+}{\it D_{uu}}}{{\,c\it R_+}\Delta_{uu}+{\,{\it k_0}
\it D_{uu}}}} \right) ^{2}{{\it \tilde{B}}}^{4}-2\,{c}^{2}{{\it D_{uu}}}^{2
}{{\it R_{uu}}}^{2}{{\it Rs}}^{2}.\nonumber
\end{split}
\end{equation}

The covariance matrix elements employed to calculate the logarithmic negativity, using the wavefunction in equation (\ref{psiuu}), are written as
\begin{footnotesize}
\begin{equation}
\langle x_1^2\rangle_{\Psi_{uu}} = \langle x_2^2 \rangle_{\Psi_{uu}} = \frac{1}{4}\big[B^2 +
\tilde{B}^2 + D_{uu}^2\big],
\end{equation}

\begin{equation}
\langle x_1x_2\rangle_{\Psi_{uu}}  =\langle x_2x_1\rangle_{\Psi_{uu}}  =\frac{1}{4}\big[B^2
-\tilde{B}^2 + D_{uu}^2\big],
\end{equation}

\begin{equation}
\begin{split}
\langle p_1^2 \rangle_{\Psi_{uu}}  = \frac{\hbar^2}{4}\bigg[
\frac{1}{B^2}+\frac{1}{\tilde{B}^2}+\frac{k_0^2}{c^2}
\bigg(\frac{B^2}{R_+^2}+\frac{\tilde{B}^2}{R_-^2}\bigg)+
\bigg(\Delta_{uu}+\frac{k_0 D_{uu}}{c R_+}\bigg)^2 \bigg],
\end{split}
\end{equation}

\begin{equation}
\begin{split}
\langle p_1p_2\rangle_{\Psi_{uu}}  = &\frac{\hbar^2}{4}\bigg[
\frac{1}{B^2}-\frac{1}{\tilde{B}^2}+\frac{k_0^2}{c^2}
\bigg(\frac{B^2} {R_{+}^2}-\frac{\tilde{B}^2}{R_{-}^2}\bigg)+
\bigg(\Delta_{uu}+\frac{k_0 D_{uu}}{c R_+}\bigg)^2\bigg],
\end{split}
\end{equation}

\begin{equation}
\langle x_1 p_2 \rangle_{\Psi_{uu}}  = \frac{\hbar}{4}\bigg[
\frac{k_0}{c}\bigg(\frac{B^2}{R_+}-\frac{\tilde{B}^2}{R_-}+\frac{D_{uu}^2}{R_+}\bigg)+D_{uu}
\Delta_{uu} \bigg],
\end{equation}
and
\begin{eqnarray}
\frac{\langle x_1p_1 +p_1x_1 \rangle_{\Psi_{uu}} }{2}
&=&\frac{\langle x_2p_2 +p_2x_2 \rangle_{\Psi_{uu}} }{2}\nonumber\\
&=&\frac{\hbar}{4}\bigg[
\frac{k_0}{c}\bigg(\frac{B^2}{R_+}+\frac{\tilde{B}^2}{R_-}+\frac{D_{uu}^2}{R_+}\bigg)+D_{uu}\Delta_{uu}
\bigg].\nonumber\\
\end{eqnarray}
\end{footnotesize}

\end{document}